\documentclass[twocolumn,prd,aps,preprintnumbers,tightenlines,showpacs,nofootinbib,eqsecnum,amsfonts,amsmath, floatfix,groupedaddress
]{revtex4-2}

\usepackage{graphicx} 
\usepackage[usenames,dvipsnames]{xcolor}
\usepackage{xspace} 
\usepackage{bm} 
\usepackage{enumerate}
\usepackage[utf8]{inputenc}
\usepackage[normalem]{ulem}
\usepackage{graphicx}
\usepackage{lineno}
\usepackage{mathrsfs,amsmath,amsfonts,amsthm,bm}

\xdefinecolor{mylinkcolor}{rgb}{0,0,0.5}
\usepackage[
	bookmarksnumbered, bookmarksopen, bookmarksopenlevel=2,
	breaklinks=true,
	colorlinks=true, filecolor=mylinkcolor, citecolor=mylinkcolor,
	linkcolor=mylinkcolor, urlcolor=mylinkcolor, menucolor=mylinkcolor,
]{hyperref}
\usepackage{orcidlink}

\newcommand{\mr}{\mathrm}
\newcommand{\ms}{\mathsf}

\DeclareMathOperator{\Order}{\mathcal{O}}

\allowdisplaybreaks

\begin{document}
\title{
Gravitational spin-orbit dynamics at the fifth-and-a-half post-Newtonian order
}
\author{Mohammed Khalil\,\orcidlink{0000-0002-6398-4428}}
\email{mohammed.khalil@aei.mpg.de}
\affiliation{Max Planck Institute for Gravitational Physics (Albert Einstein Institute), Am M\"uhlenberg 1, Potsdam 14476, Germany\\
Department of Physics, University of Maryland, College Park, MD 20742, USA}

\begin{abstract}
Accurate waveform models are crucial for gravitational-wave data analysis, and since spin has a significant effect on the binary dynamics, it is important to improve the spin description in these models. In this paper, we derive the spin-orbit (SO) coupling at the fifth-and-a-half post-Newtonian (5.5PN) order. The method we use splits the conservative dynamics into local and nonlocal-in-time parts, then relates the local-in-time part to gravitational self-force results by exploiting the simple mass-ratio dependence of the post-Minkowskian expansion of the scattering angle. We calculate the nonlocal contribution to the 5.5PN SO dynamics to eighth order in the small-eccentricity expansion for bound orbits, and to leading order in the large-eccentricity expansion for unbound orbits. For the local contribution, we obtain all the 5.5PN SO coefficients from first-order self-force results for the redshift and spin-precession invariants, except for one unknown that could be fixed in the future by second-order self-force results. However, by incorporating our 5.5PN results in the effective-one-body formalism and comparing its binding energy to numerical relativity, we find that the remaining unknown has a small effect on the SO dynamics, demonstrating an improvement in accuracy at that order.
\end{abstract}

\maketitle

\section{Introduction}

Gravitational-wave (GW) observations~\cite{LIGOScientific:2016aoc,LIGOScientific:2020ibl, LIGOScientific:2018mvr} have improved our understanding of compact binary systems, their properties, and their formation channels~\cite{LIGOScientific:2018jsj,LIGOScientific:2020kqk}.
A crucial component in searching for GW signals and inferring their parameters is accurate analytical waveform models, in which spin is an important ingredient given its significant effect on the orbital dynamics.

Three main analytical approximation methods exist for describing the dynamics during the inspiral phase:
the post-Newtonian (PN), the post-Minkowskian (PM), and the small-mass-ratio (gravitational self-force (GSF)) approximations.

The PN approximation is valid for small velocities and weak gravitational potential $v^2/c^2 \sim GM/rc^2 \ll 1$, and is most applicable for comparable-mass binaries in bound orbits. 
Many studies have contributed to improving the description of the conservative PN dynamics, for nonspinning binaries~\cite{Blanchet:2000ub,Jaranowski:1997ky,Pati:2000vt,Damour:2014jta,Damour:2015isa,Damour:2016abl,Bernard:2017ktp,Marchand:2017pir,Foffa:2019rdf,Foffa:2019yfl,Blumlein:2020pog,Blumlein:2020pyo,Larrouturou:2021dma,Blumlein:2021txe,Almeida:2021xwn}, at spin-orbit (SO) level~\cite{tulczyjew1959equations,Tagoshi:2000zg,Porto:2005ac,Faye:2006gx,Blanchet:2006gy,Damour:2007nc,Blanchet:2011zv,Hartung:2011te,Perrodin:2010dy,Porto:2010tr,Hartung:2013dza,Marsat:2012fn, Bohe:2012mr,Levi:2015uxa,Levi:2020kvb}, spin-spin~\cite{Hergt:2008jn,Porto:2006bt,Porto:2008jj,Porto:2008tb,Hergt:2010pa,Hartung:2011ea,Levi:2011eq,Levi:2015ixa,Cho:2021mqw}, and higher orders in spin~\cite{Levi:2014gsa,Levi:2016ofk,Levi:2019kgk,Levi:2020lfn,Vines:2016qwa,Siemonsen:2019dsu}.
For reviews, see Refs.~\cite{Futamase:2007zz,Blanchet:2013haa,Schafer:2018kuf,Levi:2015msa,Porto:2016pyg,Levi:2018nxp}.

The PM approximation is valid for arbitrary velocities in the weak field $GM/rc^2 \ll 1$, and is most applicable for scattering motion since relativistic velocities can be achieved.
It was pioneered by the classic results of Westpfahl~\cite{Westpfahl:1979gu,Westpfahl:1980mk}, 
with rapid progress using 
classical methods~\cite{Bel:1981be,schafer1986adm,Ledvinka:2008tk,Damour:2016gwp,Damour:2017zjx,Damour:2019lcq,Blanchet:2018yvb}, 
scattering amplitudes~\cite{Arkani-Hamed:2017jhn,Bjerrum-Bohr:2018xdl,Kosower:2018adc,Cheung:2018wkq,Bautista:2019tdr,Bern:2019nnu,Bern:2021dqo,Bjerrum-Bohr:2019kec,Bjerrum-Bohr:2021din,Cristofoli:2020uzm},
effective field theory~\cite{Foffa:2013gja,Kalin:2019rwq,Kalin:2020fhe,Kalin:2020mvi,Dlapa:2021npj},
and worldline quantum field theory~\cite{Mogull:2020sak,Jakobsen:2021smu}.
Spin effects were included in PM expansions using all these approaches in Refs. \cite{Bini:2017xzy,Bini:2018ywr,Vines:2017hyw,Vines:2018gqi,Guevara:2019fsj,Kalin:2019inp,Bern:2020buy,Chung:2020rrz,Maybee:2019jus,Bautista:2021wfy,Kosmopoulos:2021zoq,Liu:2021zxr,Jakobsen:2021lvp,Jakobsen:2021zvh}, 
and radiative contributions in Refs.~\cite{Damour:2020tta,Bini:2021gat,Bini:2021qvf,Bini:2021jmj,Saketh:2021sri}. 

The small-mass-ratio approximation $m_1/m_2 \ll 1$ is based on GSF theory, and is most applicable for extreme-mass-ratio inspirals (see, e.g., Refs.~\cite{Mino:1996nk,Quinn:1996am,Barack:2001gx,Barack:2002mh,Detweiler:2002mi,Barack:2002bt,Detweiler:2005kq,Rosenthal:2006iy,Hinderer:2008dm,Gralla:2008fg,Shah:2010bi,Keidl:2010pm,Pound:2010pj,Barack:2011ed,Pound:2012nt,Pound:2012dk,Gralla:2012db,Pound:2014koa,Pound:2017psq,vandeMeent:2016pee,vandeMeent:2017bcc} and the reviews~\cite{Barack:2009ux,Pound:2015tma,Barack:2018yvs,Pound:2021qin}.)
Analytic GSF calculations to high PN orders were performed at first order in the mass ratio for the gauge-invariant  redshift~\cite{Detweiler:2008ft,Bini:2013rfa,Kavanagh:2015lva,Hopper:2015icj,Bini:2015bfb,Kavanagh:2016idg,Bini:2018zde,Bini:2019lcd,Bini:2020zqy} and the spin-precession frequency~\cite{Dolan:2013roa,Bini:2014ica,Akcay:2016dku,Kavanagh:2017wot,Bini:2018ylh,Bini:2019lkm}.
There has also been recent important work on numerically calculating the binding energy and energy flux at second order in the mass ratio~\cite{Pound:2019lzj,Warburton:2021kwk}.

The effective-one-body (EOB) formalism~\cite{Buonanno:1998gg,Buonanno:2000ef} combines information from different analytical approximations with numerical relativity (NR) results, while recovering the strong-field test-body limit, thereby extending each approximation's domain of validity and improving the inspiral-merger-ringdown waveforms.
EOB models have been constructed for nonspinning \cite{Damour:2015isa,Damour:2000we,Buonanno:2007pf,Damour:2008gu,Pan:2011gk,Nagar:2019wds}, spinning~\cite{Damour:2008qf,Barausse:2009xi,Barausse:2011ys,Nagar:2011fx,Damour:2014sva,Balmelli:2015zsa,Khalil:2020mmr,Pan:2013rra,Taracchini:2012ig,Taracchini:2013rva,Babak:2016tgq,Bohe:2016gbl,Cotesta:2018fcv,Ossokine:2020kjp,Nagar:2018plt,Nagar:2018zoe}, and eccentric binaries~\cite{Bini:2012ji,Hinderer:2017jcs,Nagar:2021gss,Khalil:2021txt}.
In addition, information from the post-Minkowskian ~\cite{Damour:2016gwp,Damour:2017zjx,Antonelli:2019ytb,Damgaard:2021rnk} and small mass-ratio approximations~\cite{Damour:2009sm,Barausse:2011dq,Akcay:2012ea,Antonelli:2019fmq} have been incorporated in EOB models.

Recently, a method~\cite{Bini:2019nra}, sometimes dubbed the ``Tutti Frutti'' method~\cite{Bini:2021gat}, that combines all these formalisms has been used to derive PN results valid for arbitrary mass ratios from GSF results at first order in the mass ratio. 
The method relies on the simple mass-ratio dependence of the PM-expanded scattering angle~\cite{Damour:2019lcq} (see also Ref.~\cite{Vines:2018gqi}), making it possible to relate the local-in-time part of the Hamiltonian, or radial action, to GSF invariants, such as the redshift and precession frequency.
The nonlocal-in-time part of the conservative dynamics, due to backscattered radiation emitted at earlier times, is derived separately, since it is calculated in an eccentricity expansion that differs between bound and unbound orbits.
This approach has been used to derive the 5PN conservative dynamics for nonspinning binaries except for two coefficients~\cite{Bini:2019nra,Bini:2020wpo}, the 6PN dynamics except for four coefficients~\cite{Bini:2020hmy,Bini:2020nsb}, and the full 4.5PN SO and 5PN aligned spin$_1$-spin$_2$ dynamics~\cite{Antonelli:2020aeb,Antonelli:2020ybz}.

In this paper, we determine the 5.5PN SO coupling for the two-body dynamics, which is the fourth-subleading PN order, except for one coefficient at second order in the mass ratio.
Throughout, we perform all calculations for spins aligned, or antialigned, with the direction of the orbital angular momentum. However, the results are valid for precessing spins~\cite{Antonelli:2020aeb}, since at SO level, the spin vector only couples to the angular momentum vector.

The results of this paper and the procedure used can be summarized as follows:
\begin{enumerate}
\item In Sec.~\ref{sec:nonloc}, we calculate the nonlocal contribution to the 5.5PN SO Hamiltonian for bound orbits, in a small-eccentricity expansion up to eighth order in eccentricity.
We do this for a harmonic-coordinates Hamiltonian, then incorporate those results into the gyro-gravitomagnetic factors in an EOB Hamiltonian.

\item In Sec.~\ref{sec:local}, we determine the local contribution by relating the coefficients of the local Hamiltonian to those of the PM-expanded scattering angle.
We then calculate the redshift and spin-precession invariants from the total Hamiltonian, and match their small-mass-ratio expansion to first-order self-force (1SF) results.
This allows us to recover all the coefficients of the local part except for one unknown. However, by computing the EOB binding energy and comparing it to NR, we show that the effect of the remaining unknown on the dynamics is small.

\item In Sec.~\ref{sec:nonlocscatter}, we complement our results for unbound orbits by calculating the nonlocal part of the gauge-invariant scattering angle, to leading order in the large-eccentricity expansion.

\item In Sec.~\ref{sec:radAction}, we provide two gauge-invariant quantities that characterize bound orbits: the radial action as a function of energy and angular momentum, and the circular-orbit binding energy as a function of frequency.
\end{enumerate}
We conclude in  Sec.~\ref{sec:conc} with a discussion of the results, and provide in Appendix~\ref{app:qkepler} a summary of the quasi-Keplerian parametrization at leading SO order.
The main results of this paper are provided in the  Supplemental Material as a \textsc{Mathematica} file~\cite{ancprd}.

\subsection*{Notation}
We use the metric signature $(-,+,+,+)$, and units in which $G=c=1$, but sometimes write them explicitly in PM and PN expansions for clarity. 

For a binary with masses $m_1$ and $m_2$, with $m_2 \geq m_1$, and spins $\bm{S}_1$ and $\bm{S}_2$, we define the following combinations of the masses:
\begin{gather}\label{massmap}
M= m_1 + m_2, \quad \mu = \frac{m_1m_2}{M}, \quad \nu = \frac{\mu}{M}, \nonumber\\
q = \frac{m_1}{m_2}, \quad \delta =\frac{m_2 - m_1}{M},
\end{gather}
define the mass-rescaled spins
\begin{equation}
\bm{a}_1 = \frac{\bm{S}_1}{m_1}, \qquad
\bm{a}_2 = \frac{\bm{S}_2}{m_2},
\end{equation}
the dimensionless spin magnitudes
\begin{gather}
\chi_1 = \frac{|\bm{S}_1|}{m_1^2}, \qquad 
\chi_2 = \frac{|\bm{S}_2|}{m_2^2},
\end{gather}
and the spin combinations
\begin{gather}
\bm{S} = \bm{S}_1 + \bm{S}_2, \qquad \bm{S}^* = \frac{m_2}{m_1} \bm{S}_1 + \frac{m_1}{m_2} \bm{S}_2, \nonumber\\
\chi_S = \frac{1}{2}(\chi_1 + \chi_2), \qquad
\chi_A = \frac{1}{2}(\chi_2 - \chi_1).
\end{gather}

We use several variables related to the total energy $E$ of the binary system: the binding energy $\bar{E}= E - M c^2$, the mass-rescaled energy $\Gamma = E / M$, and the effective energy $E_\text{eff}$ defined by the energy map 
\begin{equation}
E = M \sqrt{1 + 2\nu \left(\frac{E_\text{eff}}{\mu} - 1\right)}\,.
\end{equation}
We also define the asymptotic relative velocity $v$ and Lorentz factor $\gamma$ via
\begin{gather}
\gamma = \frac{E_\text{eff}}{\mu}, \nonumber\\
v = \frac{\sqrt{\gamma^2 - 1}}{\gamma}, \quad \leftrightarrow \quad \gamma = \frac{1}{\sqrt{1-v^2}},
\end{gather} 
and define the dimensionless energy variable
\begin{equation}
\varepsilon = \gamma^2 - 1 = \gamma^2 v^2.
\end{equation}
(Note that $\varepsilon$ used here is denoted $p_\infty^2$ in Ref.~\cite{Bini:2020wpo}.)

The magnitude of the orbital angular momentum is denoted $L$, and is related to the relative position $r$, radial momentum $p_r$, and total linear momentum $p$ via 
\begin{equation}
p^2 = p_r^2 + \frac{L^2}{r^2}.
\end{equation}

We often use dimensionless rescaled quantities, such as
\begin{gather}
r=\frac{r^\text{phys}}{M}, \quad L=\frac{L^\text{phys}}{M\mu}, \quad p_r = \frac{p_r^\text{phys}}{\mu}, \nonumber\\
E = \frac{E^\text{phys}}{\mu}, \quad  H = \frac{H^\text{phys}}{\mu},
\end{gather}
and similarly for related variables, e.g. $\bar{E}=\bar{E}^\text{phys}/\mu$, etc. 
It should be clear from the context whether physical or rescaled quantities are being used.

\section{Nonlocal 5.5PN SO Hamiltonian for bound orbits}
\label{sec:nonloc}
The total conservative action at a given PN order can be split into local and nonlocal-in-time parts, such that
\begin{equation}
S_\text{tot}^\text{nPN} = S_\text{loc}^\text{nPN} + S_\text{nonloc}^\text{nPN},
\end{equation}
where the nonlocal part is due to tail effects projected on the conservative dynamics~\cite{Blanchet:1987wq,Damour:2014jta,Bernard:2015njp}, i.e., radiation emitted at earlier times and backscattered onto the binary.
The nonlocal contribution starts at 4PN order, and has been derived for nonspinning binaries up to 6PN order~\cite{Damour:2015isa,Bini:2020wpo,Bini:2020hmy}.
In this section, we derive the leading-order spin contribution to the nonlocal part, which is at 5.5PN order.

The nonlocal part of the action can be calculated via the following integral:
\begin{equation}
S_\text{nonloc}^\text{LO} = \frac{GM}{c^3} \int dt\, \text{Pf}_{2s/c} \int \frac{dt'}{|t-t'|} \mathcal{F}_\text{LO}^\text{split}(t,t'),
\end{equation}
where the label `LO' here means that we include the \emph{leading-order} nonspinning and SO contributions, 
and where the Hadamard \textit{partie finie} (Pf) operation is used since the integral is logarithmically divergent at $t'= t$.
The time-split (or time-symmetric) GW energy flux $\mathcal{F}_\text{LO}^\text{split}(t,t')$ is written in terms of the source multipole moments as~\cite{Damour:2014jta}
\begin{equation}
\label{Fsplit}
\mathcal{F}_\text{LO}^\text{split}(t,t') = \frac{G}{c^5} \left[
\frac{1}{5} I_{ij}^{(3)}(t)I_{ij}^{(3)}(t') + \frac{16}{45c^2} J_{ij}^{(3)}(t)J_{ij}^{(3)}(t')
\right].
\end{equation}
The mass quadrupole $I^{ij}$ and the current quadrupole $J^{ij}$ (in harmonic coordinates and using the Newton-Wigner spin-supplementary condition~\cite{pryce1948mass,newton1949localized}) are given by \cite{blanchet1989post,Kidder:1995zr}
\begin{align}
\label{sourceMom}
I_{ij} &= m_1 x_1^{\langle i} x_1^{j \rangle} + \frac{3}{c^3} x_1^{\langle i} (\bm{v}_1\times \bm{S}_1)^{j \rangle}  \nonumber\\
&\quad  - \frac{4}{3c^3} \frac{d}{dt} x_1^{\langle i}(\bm{x}_1\times \bm{S}_1)^{j \rangle} + 1 \leftrightarrow 2, \nonumber\\
J_{ij} &= m_1 x_1^{\langle i} (\bm{x}_1 \times \bm{v}_1)^{j \rangle} + \frac{3}{2c} x_1^{\langle i} S_1^{j \rangle} + 1 \leftrightarrow 2,
\end{align}
where the indices in angle brackets denote a symmetric trace-free part. 

As was shown in Refs.~\cite{Damour:2014jta,Damour:2015isa}, the nonlocal part of the action can be written in terms of $\tau \equiv t' - t$ as
\begin{align}
\label{nonlocH}
S_\text{nonloc}^\text{LO} &= - \int dt\, \delta H_\text{nonloc}^\text{LO}(t), \nonumber\\
\delta H_\text{nonloc}^\text{LO}(t) &= -\frac{GM}{c^3} \text{Pf}_{2s/c} \int \frac{d\tau}{|\tau|} \mathcal{F}_\text{LO}^\text{split}(t,t+\tau) \nonumber\\
&\quad 
+ 2\frac{GM}{c^3} \mathcal{F}_\text{LO}^\text{split}(t,t) \ln\left(\frac{r}{s}\right).
\end{align}
Following Ref.~\cite{Bini:2020wpo}, we choose the arbitrary length scale $s$ entering the \textit{partie finie} operation to be the radial distance $r$ between the two bodies in harmonic coordinates. 
This has the advantage of simplifying the local part by removing its dependence on $\ln r$.

\subsection{Computation of the nonlocal Hamiltonian in a small-eccentricity expansion}
The integral for the nonlocal Hamiltonian in Eq.~\eqref{nonlocH} can be performed in a small-eccentricity expansion using the quasi-Keplerian parametrization~\cite{damour1985general}, which can be expressed, up to 1.5PN order, by the following equations:
\begin{align}
r &= a_r (1 - e_r \cos u), \label{qKr}\\
\ell &= n t = u - e_t \sin u, \label{kepEq}\\
\phi &= 2 K \arctan\left[\sqrt{\frac{1+e_\phi}{1-e_\phi}} \tan \frac{u}{2}\right], \label{qKphi}
\end{align}
where $a_r$ is the semi-major axis, $u$ the eccentric anomaly, $\ell$ the mean anomaly, $n$ the mean motion (radial angular frequency), $K$ the periastron advance, and ($e_r,e_t,e_\phi$) the radial, time, and phase eccentricities.

The quasi-Keplerian parametrization was generalized to 3PN order in Ref.~\cite{Memmesheimer:2004cv}, and including SO and spin-spin contributions in Refs.~\cite{Tessmer:2010hp,Tessmer:2012xr}.  
We summarize in Appendix~\ref{app:qKepBound} the relations between the quantities used in the quasi-Keplerian parametrization and the energy and angular momentum at leading SO order.
Using the quasi-Keplerian parametrization, we express the source multipole moments in terms of the variables ($a_r,e_t,t$), and expand the moments in eccentricity.

In the center-of-mass frame, the position vectors of the two bodies, $\bm{x}_1$ and $\bm{x}_2$, are related to $\bm{x}\equiv\bm{x}_1-\bm{x}_2$ via~\cite{Will:2005sn}
\begin{align}
x_1^i &= \frac{m_2}{M} x^i - \frac{\nu}{2c^3\delta M} {\epsilon^i}_{jk} v^j (S^k - {S^*}^k), \nonumber\\
x_2^i &= -\frac{m_1}{M} x^i - \frac{\nu}{2c^3\delta M} {\epsilon^i}_{jk} v^j (S^k - {S^*}^k),
\end{align}
where $\bm{v}\equiv \bm{v}_1 - \bm{v}_2$, and hence the source moments from Eq.~\eqref{sourceMom} can be written as
\begin{align}
I_{ij} &= M \nu x^{\langle i} x^{j \rangle} 
+ \frac{1}{3c^3} \bigg[
\frac{m_2^2}{M^2} \Big(4 v^{\langle i} (\bm{S}_1\times \bm{x})^{j \rangle} \nonumber\\
&\quad\qquad
- 5 x^{\langle i} (\bm{S}_1\times \bm{v})^{j \rangle}\Big) 
+ 1\leftrightarrow 2
\bigg], \\
J_{ij} &= M \delta \nu   x^{\langle i} (\bm{v}\times \bm{x})^{j \rangle}
+ \frac{3}{2c} \left[\frac{m_2}{M} x^{\langle i} S_1^{j \rangle}
- \frac{m_1}{M} x^{\langle i} S_2^{j \rangle} \right].\nonumber
\end{align}
In polar coordinates,
\begin{align}
\bm{x} &= r (\cos \phi,\, \sin \phi), \nonumber\\
\bm{v} &= \dot{r} (\cos \phi,\, \sin \phi) + r \dot{\phi} (-\sin\phi,\, \cos\phi),
\end{align}
with $r$ and $\phi$ given by Eqs.~\eqref{qKr} and~\eqref{qKphi}, $e_r$ and $e_\phi$ related to $e_t$ via Eqs.~\eqref{eret} and ~\eqref{ephiet},
 while $\dot{r}$ and $\dot{\phi}$ are given by
\begin{align}
\dot{r} &= \frac{e_t \sin u}{\sqrt{a_r} \left(1-e_t \cos u\right)}\,, \nonumber\\
\dot{\phi} &=\frac{\sqrt{a_r-a_r e_t^2}}{a_r^2 \left(1-e_t \cos u\right)^2}\,,
\end{align}
which are only needed at leading order.
We then write the eccentric anomaly $u$ in terms of time $t$ using Kepler's equation~\eqref{kepEq}, which has a solution in terms of a Fourier-Bessel series expansion
\begin{align}
u &= n t + \sum_{k=1}^{\infty} \frac{2}{k} J_k(ke_t) \sin(kn t) \nonumber\\
&\simeq
n t + e_t \sin(n t) + \frac{1}{2} e_t^2 \sin(2 n t) + \dots.
\end{align}
 
We perform the eccentricity expansion for the nonlocal part up to $\Order(e_t^8)$ since it corresponds to an expansion to $\Order(p_r^8)$, which is the highest power of $p_r$ in the 5.5PN SO local part. 
However, to simplify the presentation, we write the intermediate steps only expanded to $\Order(e_t)$.

Plugging the expressions for $(r,\phi,\dot{r},\dot{\phi})$ in terms of $(a_r,e_t,t)$ into the source moments used in the time-split energy flux~\eqref{Fsplit} and expanding in eccentricity yields
\begin{widetext}
\begin{align}
&\mathcal{F}_\text{LO}^\text{split}(t,t+\tau) = \nu^2 a_r^4 n^6 \left\lbrace \frac{32}{5} \cos (2 n \tau )
+\frac{12}{5} e_t \left[9 \cos (nt+3 n\tau)+9 \cos (nt-2n \tau)-\cos (n t-n\tau )-\cos (n t+2n \tau )\right] \right\rbrace \nonumber\\
&\quad
+ \frac{8 \nu^2}{15} n^6 a_r^{5/2} \bigg\lbrace
48 n \tau  \sin (2 n \tau ) \left(2 \delta  \chi _A-\nu  \chi _S+2 \chi _S\right)
-32 \cos (2 n \tau ) \left(9 \delta  \chi _A-5 \nu  \chi _S+9 \chi _S\right)
-\cos (n \tau ) \left(\delta  \chi _A-4 \nu  \chi _S+\chi _S\right) \nonumber\\
&\qquad
+e_t \cos \left(n t+\frac{n \tau }{2}\right) \bigg[
\cos \left(\frac{3 n \tau }{2}\right) \left(352 \delta  \chi _A+352\chi _S-157 \nu \chi _S\right)
-27 \cos \left(\frac{5 n \tau }{2}\right) \left(84 \delta  \chi _A-47 \nu  \chi _S+84 \chi _S\right) \nonumber\\
&\qquad\qquad
-36 n \tau  \left(\sin \left(\frac{3 n \tau }{2}\right)-9 \sin \left(\frac{5 n \tau }{2}\right)\right) \left(2 \delta  \chi _A-\nu  \chi _S+2 \chi _S\right)
\bigg]
\bigg\rbrace + \Order\left(e_t^2\right),
\end{align}
the orbit average of which is given by
\begin{align}
\left\langle \mathcal{F}_\text{LO}^\text{split}(t,t+\tau)\right\rangle &= \frac{n}{2\pi} \int_{0}^{2\pi/n} \mathcal{F}_\text{LO}^\text{split}(t,t+\tau) dt \nonumber\\
&=\frac{32}{5} \nu^2 n^6 a_r^4 \cos (2 n \tau ) + \frac{8}{15} \nu^2 n^6 a_r^{5/2} \Big[ 
48 n \tau  \sin (2 n \tau ) \left(2 \delta  \chi _A-\nu  \chi _S+2 \chi _S\right) -\cos (n \tau ) \left(\delta  \chi _A-4 \nu  \chi _S+\chi _S\right)\nonumber\\
&\quad\qquad
-32 \cos (2 n \tau ) \left(9 \delta  \chi _A-5 \nu  \chi _S+9 \chi _S\right)  
\Big] + \Order\left(e_t^2\right). 
\end{align}
In the limit $\tau = 0$, this equation agrees with the eccentricity expansion of the energy flux from Eq.~(64) of Ref.~\cite{Tessmer:2012xr}.

Then, we perform the \textit{partie finie} operation with time scale $2s/c$ using Eq.~(4.2) of Ref.~\cite{Damour:2014jta}, which reads
\begin{equation}
\text{Pf}_T \int_{0}^{\infty} \frac{dv}{v} g(v) = \int_{0}^{T} \frac{dv}{v} [g(v) - g(0)] + \int_{T}^{\infty} \frac{dv}{v} g(v).
\end{equation}
The first line of Eq.~\eqref{nonlocH} yields
\begin{align}
&- \text{Pf}_{2s/c} \int \frac{d\tau}{|\tau|} \left\langle \mathcal{F}_\text{LO}^\text{split}(t,t+\tau)\right\rangle =
\frac{64}{5} \nu^2 n^6 a_r^4 \left[\ln (4 n s)+\gamma_E\right] 
-\frac{16}{15} \nu^2 n^6 a_r^{5/2} \Big\lbrace
\left[289 \delta  \chi _A+ (289-164 \nu)\chi _S\right]\ln (n s)
 \nonumber\\
&\qquad\qquad
+\left[289 \gamma_E +48+577 \ln 2\right]\delta \chi _A +\left[\gamma_E  (289-164 \nu )-12 \nu  (2+27 \ln 2)+48+577 \ln 2\right] \chi _S
\Big\rbrace + \Order\left(e_t^2\right),
\end{align}
while the second line 
\begin{align}
2 \left\langle \mathcal{F}_\text{LO}^\text{split}(t,t)\right\rangle \ln\left(\frac{r}{s}\right) =
\frac{64}{5} \nu^2 n^6 a_r^4 \ln \left(\frac{a_r}{s}\right)-\frac{16}{15} \nu^2 n^6 a_r^{5/2} \ln \left(\frac{a_r}{s}\right) \left[289 \delta  \chi _A+(289-164 \nu ) \chi _S\right] +  \Order\left(e_t^2\right).
\end{align}
Adding the two expressions removes the dependence on $s$.

When performing the calculation to $\Order\left(e_t^8\right)$, we obtain the following Delaunay-averaged nonlocal Hamiltonian:
\begin{align}
\label{Hnonloc}
\left\langle \delta H_\text{nonloc}^\text{LO}\right\rangle  &=
\frac{\nu^2}{a_r^5} \left[\mathcal{A}^\text{4PN}(e_t) + \mathcal{B}^\text{4PN}(e_t) \ln a_r\right] \nonumber\\
&\quad
+ \frac{\nu^2\delta\chi_A}{a_r^{13/2}} 
\Bigg\lbrace
\frac{584}{15} \ln a_r-\frac{64}{5}-\frac{464}{3} \ln 2 -\frac{1168}{15} \gamma_E \nonumber\\
&\quad\quad
+ e_t^2 \left[\frac{2908}{5} \ln a_r-\frac{5816}{5}\gamma_E+\frac{2172}{5}-\frac{3304}{15} \ln 2-\frac{10206}{5} \ln 3\right] \nonumber\\
&\quad\quad
+e_t^4 \left[\frac{43843}{15} \ln a_r-\frac{87686}{15} \gamma_E+\frac{114991}{30}-\frac{201362}{5} \ln 2+\frac{48843}{4} \ln 3\right] \nonumber\\
&\quad\quad
+e_t^6 \left[\frac{55313}{6} \ln a_r-\frac{55313}{3} \gamma_E+\frac{961807}{60}+\frac{6896921}{45} \ln 2-\frac{3236031}{160} \ln 3-\frac{24296875}{288} \ln 5\right] \nonumber\\
&\quad\quad
+e_t^8 \left[\frac{134921}{6} \ln a_r-\frac{134921}{3} \gamma_E+\frac{135264629}{2880}-\frac{94244416}{135}  \ln 2+\frac{12145234375}{27648} \ln 5-\frac{1684497627}{5120} \ln 3\right]
\!\Bigg\rbrace \nonumber\\
&\quad
+\frac{\nu^2 \chi_S}{a_r^{13/2}} \Bigg\lbrace
-\frac{64}{5}+\frac{32 \nu }{5}+\left(\frac{896 \nu }{15}-\frac{1168}{15}\right) \gamma_E+\left(\frac{576 \nu }{5}-\frac{464}{3}\right) \ln 2 + \left(\frac{584}{15}-\frac{448 \nu }{15}\right) \ln a_r \nonumber\\
&\quad\quad
+e_t^2 \bigg[\frac{2172}{5}-\frac{4412 \nu }{15} +\left(\frac{4216 \nu }{5}-\frac{5816}{5}\right) \gamma_E+\left(\frac{5192 \nu }{15}-\frac{3304}{15}\right) \ln 2 +\left(\frac{6561 \nu }{5}-\frac{10206}{5}\right) \ln 3\nonumber\\
&\quad\qquad 
+ \left(\frac{2908}{5}-\frac{2108 \nu }{5}\right) \ln a_r\bigg] 
+e_t^4 \bigg[\frac{114991}{30}-\frac{38702 \nu }{15}+\left(\frac{62134 \nu }{15}-\frac{87686}{15}\right) \gamma_E \nonumber\\
&\quad\qquad
+\left(\frac{386414 \nu }{15}-\frac{201362}{5}\right) \ln 2 +\left(\frac{48843}{4}-\frac{28431 \nu }{4}\right) \ln 3+\left(\frac{43843}{15}-\frac{31067 \nu }{15}\right) \ln a_r\bigg]  \nonumber\\
&\quad\quad
+e_t^6 \bigg[
\frac{961807}{60}-\frac{215703 \nu }{20} + \left(\frac{193718 \nu }{15}-\frac{55313}{3}\right) \gamma_E +\left(\frac{6896921}{45}-\frac{12343118 \nu }{135}\right) \ln 2 \nonumber\\
&\quad\qquad
+\left(\frac{3768201 \nu }{320}-\frac{3236031}{160}\right) \ln 3+\left(\frac{92421875 \nu }{1728}-\frac{24296875}{288}\right) \ln 5 + \left(\frac{55313}{6}-\frac{96859 \nu }{15}\right) \ln a_r
\bigg] \nonumber\\
&\quad\quad
+e_t^8 \bigg[\frac{135264629}{2880}-\frac{45491177 \nu }{1440}+\left(\frac{93850 \nu }{3}-\frac{134921}{3}\right) \gamma_E +\left(\frac{118966123 \nu }{270}-\frac{94244416}{135}\right) \ln 2 \nonumber\\
&\quad\qquad
+\left(\frac{537837489 \nu }{2560}-\frac{1684497627}{5120}\right) \ln 3 +\left(\frac{12145234375}{27648}-\frac{3790703125 \nu }{13824}\right) \ln 5 \nonumber\\
&\quad\qquad
+ \left(\frac{134921}{6}-\frac{46925 \nu }{3}\right) \ln a_r\bigg]
\Bigg\rbrace + \Order\left(e_t^{10}\right),
\end{align}
\end{widetext}
where the functions $\mathcal{A}^\text{4PN}(e_t)$ and $\mathcal{B}^\text{4PN}(e_t)$ in the 4PN part are given in Table I of Ref.~\cite{Bini:2020wpo}.

\subsection{Nonlocal part of the EOB Hamiltonian}
The (dimensionless) EOB Hamiltonian is given by the energy map
\begin{equation}
\label{Heob}
H_\text{EOB} = \frac{1}{\nu} \sqrt{1 + 2 \nu \left(H_\text{eff} - 1\right)}\,,
\end{equation}
where the effective Hamiltonian
\begin{align}
H_\text{eff} &= \sqrt{A(r) \left[1 + p^2 + \left(A(r) \bar{D}(r) - 1\right) p_r^2 + Q(r,p_r)\right]} \nonumber\\
&\quad 
+ \frac{1}{c^3 r^3} \bm{L}\cdot\left[g_S(r,p_r) \bm{S} + g_{S^*}(r,p_r) \bm{S}^*\right].
\end{align}

The nonspinning potentials $A,\bar{D},$ and $Q$ were obtained at 4PN order in Ref.~\cite{Damour:2015isa}.
The 4.5PN gyro-gravitomagnetic factors, $g_S$ and $g_{S^*}$, are given by Eq.~(5.6) of Ref.~\cite{Antonelli:2020ybz}, and are in a gauge such that they are independent of the angular momentum.
Note that the gyro-gravitomagnetic factors are the same for both aligned and precessing spins, since the spin vector only couples to the angular momentum vector at SO level. Hence, even though the calculations are specialized to aligned spins, the final result for the gyro-gravitomagnetic factors is valid for precessing spins.

Splitting the potentials $A,\bar{D},Q$ into a local and a nonlocal piece, and writing the gyro-gravitomagnetic factors as
\begin{align}
\label{gyros}
g_S &= 2 + \dots + \frac{1}{c^8}\left(g_{S}^\text{5.5PN,loc} +  g_S^\text{5.5PN,nonloc}\right),  \nonumber\\
g_{S^*} &= \frac{3}{2} + \dots + \frac{1}{c^8} \left(g_{S^*}^\text{5.5PN,loc} +  g_{S^*}^\text{5.5PN,nonloc}\right)
\end{align}
yields the following LO nonlocal part of the PN-expanded effective Hamiltonian 
\begin{align}
H_\text{eff} &= H_\text{eff}^\text{loc} + \frac{1}{c^8} H_\text{eff}^\text{nonloc} + \Order(\text{5PN, 6.5PN SO}), \nonumber\\
H_\text{eff}^\text{nonloc} &= \frac{1}{2} \left(A^\text{nonloc} +  \bar{D}^\text{nonloc} p_r^2 +  Q^\text{nonloc}\right) \nonumber\\
&\quad
+ \frac{\nu \bm{L}}{c^3r^3} \cdot \left[\bm{S}  g_S^\text{5.5PN,nonloc} + \bm{S}^*  g_{S^*}^\text{5.5PN,nonloc}\right].
\end{align}

Then, we write the nonlocal piece of the potentials and gyro-gravitomagnetic factors in terms of unknown coefficients, calculate the Delaunay average of $H_\text{eff}^\text{nonloc}$ in terms of the EOB coordinates $(a_r,e_t)$, and match it to the harmonic-coordinates Hamiltonian from Eq.~\eqref{Hnonloc}. Since harmonic and EOB coordinates agree at leading SO order, no canonical transformation is needed between the two at that order.

This yields the results in Table IV of Ref.~\cite{Bini:2020wpo} for the 4PN nonspinning part, and the following SO part expanded to $\Order\left(p_r^8\right)$:
\begin{widetext}
\begin{align}
\label{gyroNonloc}
&g_S^\text{5.5PN,nonloc} = 2 \nu \bigg\lbrace\!\!
\left(\frac{292 }{15}\ln r -\frac{32}{5}-\frac{584}{15}\gamma_E-\frac{232}{3} \ln 2\right) \frac{1}{r^4}
+ \left(\frac{12782}{15}-104 \gamma_E+\frac{32744}{15} \ln 2-\frac{11664 }{5}\ln 3+52 \ln r\right)  \frac{p_r^2}{r^3} \nonumber\\
&\quad\qquad
+ \left(\frac{12503}{15}-\frac{635456}{9} \ln 2+\frac{218943}{5}\ln 3 \right) \frac{p_r^4}{r^2}
+ \left(\frac{38246}{25}+\frac{176799232}{225} \ln 2-\frac{2517237}{10} \ln 3-\frac{3015625}{18} \ln 5\right) \frac{p_r^6}{r} \nonumber\\
&\quad\qquad
+ \left(\frac{503099}{350}-\frac{898982848}{189} \ln 2+\frac{6352671875}{3024} \ln 5-\frac{31129029}{400} \ln 3\right) p_r^8 + \Order\left(p_r^{10}\right)\bigg\rbrace, \nonumber\\
&g_{S^*}^\text{5.5PN,nonloc} = \frac{3}{2} \nu \bigg\lbrace\!\! 
\left(16 \ln r-\frac{32}{5}-32 \gamma_E-\frac{2912}{45}  \ln 2\right) \frac{1}{r^4}
+ \left(\frac{35024}{45}-\frac{1024 \gamma_E}{15}+\frac{93952}{45} \ln 2-\frac{10692}{5} \ln 3 +\frac{512}{15} \ln r\right) \frac{p_r^2}{r^3} \nonumber\\
&\quad\qquad
+ \left(\frac{9232}{15}-\frac{2978624}{45} \ln 2+\frac{206064}{5}\ln 3\right)\frac{ p_r^4}{r^2}
+ \left(\frac{33048}{25}+\frac{1497436672}{2025} \ln 2-\frac{1199934 }{5}\ln 3 -\frac{12593750}{81} \ln 5\right) \frac{p_r^6}{r} \nonumber\\
&\quad\qquad
+ \left(\frac{651176}{525}-\frac{9076395968}{2025} \ln 2+\frac{2226734375 }{1134}\ln 5-\frac{697653}{14} \ln 3\right) p_r^8 + \Order\left(p_r^{10}\right)\bigg\rbrace.
\end{align}
\end{widetext}

\section{Local 5.5PN SO Hamiltonian and scattering angle}
\label{sec:local}
In this section, we determine the local part of the Hamiltonian and scattering angle from 1SF results by making use of the simple mass dependence of the PM-expanded scattering angle.

\subsection{Mass dependence of the scattering angle}
Based on the structure of the PM expansion, Poincar\'e symmetry, and dimensional analysis, Ref.~\cite{Damour:2019lcq} (see also Ref.~\cite{Vines:2018gqi}) showed that the magnitude of the impulse (net change in momentum), for nonspinning systems in the center-of-mass frame, has the following dependence on the masses:
\begin{align}
\ms Q&= (\Delta p_{1\mu}\Delta p^{1\mu})^{1/2} \nonumber\\
&=\frac{2Gm_1m_2}{b}\bigg[\ms Q^\mr{1PM}
\nonumber\\
&\quad+\frac{G}{b}\bigg(m_1\ms Q^\mr{2PM}_{m_1}+m_2\ms Q^\mr{2PM}_{m_2}\bigg)
\nonumber\\
&\quad+\frac{G^2}{b^2}\bigg(m_1^2\ms Q^\mr{3PM}_{m_1^2}+m_1m_2 \ms Q_{m_1m_2}^\mr{3PM}+m_2^2\ms Q^\mr{3PM}_{m_2^2}\bigg)
\nonumber\\
&\quad+\cdots\bigg],
\end{align}
where each PM order is a homogeneous polynomial in the two masses.
For nonspinning bodies, the $\ms Q$'s on the right-hand side are functions only of energy (or velocity $v$).
This mass dependence has been extended in Ref.~\cite{Antonelli:2020ybz} to include spin, such that
\begin{align}
\ms Q^{n\mr{PM}}_{m_1^im_2^j}&=\ms Q^{n\mr{PM}}_{m_1^im_2^j}\left(v,\frac{a_1}{b},\frac{a_2}{b}\right)
\nonumber\\
&=\ms Q^{n\mr{PM}}_{m_1^im_2^ja^0}(v)
+\frac{a_1}{b}\ms Q^{n\mr{PM}}_{m_1^im_2^ja_1}(v)\nonumber\\
&\quad
+\frac{a_2}{b}\ms Q^{n\mr{PM}}_{m_1^im_2^ja_2}(v) + \Order(a_i^2),
\end{align}
where $b$ is the \emph{covariant} impact parameter defined as the orthogonal distance between the incoming worldlines when using the covariant (Tulczyjew-Dixon) spin-supplementary condition~\cite{Tulczyjew:1959,Dixon:1979} $p_{\ms i\mu}S_{\ms i}^{\mu\nu}=0$. (See Refs.~\cite{Vines:2016unv,Vines:2017hyw,Vines:2018gqi,Antonelli:2020ybz} for more details.)

The scattering angle $\chi$ by which the two bodies are deflected in the center-of-mass frame is related to $\ms Q$ via~\cite{Damour:2019lcq}
\begin{equation}
\sin\frac{\chi}{2}=\frac{\ms Q}{2P_\text{c.m.}},
\end{equation}
where $P_\text{c.m.}$ is the magnitude of the total linear momentum in the center-of-mass frame and is given by
\begin{equation}
P_\text{c.m.}=\frac{m_1m_2}{E}\sqrt{\gamma^2-1}\,,
\end{equation}
where we recall that
\begin{align}
\label{Egamma}
E^2&=m_1^2+m_2^2+2m_1m_2\gamma \nonumber\\
&= M^2 \left[1 + 2\nu (\gamma -1 )\right], \nonumber\\
\gamma &= \frac{1}{\sqrt{1 - v^2}}.
\end{align}
Therefore, the scattering angle scaled by $E/m_1m_2$ has the same mass dependence as $\ms Q$. (Equivalently, $\chi/\Gamma$ has the same mass dependence as $\ms Q/\mu$, where $\Gamma\equiv E/M$.)

For nonspinning binaries, and because of the symmetry under the exchange of the two bodies' labels, the mass dependence of $\chi/\Gamma$ can be written as a polynomial in the symmetric mass ratio $\nu$.
This is because any homogeneous polynomial in the masses $(m_1,m_2)$ of degree $n$ can be written as polynomial in $\nu$ of degree $\lfloor n/2 \rfloor$. For example,
\begin{align}
&c_1 m_1^3+c_2 m_1^2 m_2 +c_2  m_1 m_2^2+c_1 m_2^3 \nonumber\\
&\qquad\qquad
= M^3[ c_1 +  (c_2-3 c_1)\nu],
\end{align}
for some mass-independent factors $c_i$. 
Hence, at each $n$PM order, $\chi/\Gamma$ is a polynomial in $\nu$ of degree $\lfloor (n-1)/2 \rfloor$.

When including spin, we also obtain a dependence on the antisymmetric mass ratio $\delta\equiv (m_2 - m_1)/ M $, since
\begin{align}
&a_1 \left(c_1 m_1^3+ c_2 m_1^2 m_2 + c_3  m_1 m_2^2+ c_4 m_2^3\right) \nonumber\\
&\qquad\qquad
= M^3 a_1 \left(\alpha_1  +\alpha_2 \delta+\alpha_3 \nu +\alpha_4 \nu\delta\right),
\end{align}
where $\alpha_i$ are some linear combinations of $c_i$.

Thus, we find that the scattering angle, up to 5PM and to linear order in spin, has the following mass dependence:
\begin{align}
\chi &= \chi_{a^0} + \chi_{a} + \Order(a^2), \\
\frac{\chi_{a^0}}{\Gamma} &= \frac{GM}{b}\ms X_1^0 +\Big(\frac{GM}{b}\Big)^2\ms X_2^0 \nonumber\\
&\quad
+\Big(\frac{GM}{b}\Big)^3\Big[\ms X_3^0+\nu \ms X_3^{0,\nu}\Big]
\nonumber\\
&\quad 
+\Big(\frac{GM}{b}\Big)^4\Big[\ms X_4^0+\nu \ms X_4^{0,\nu}\Big] \nonumber\\
&\quad + \Big(\frac{GM}{b}\Big)^5\Big[\ms X_5^0+\nu \ms X_5^{0,\nu}+\nu^2 \ms X_5^{0,\nu^2}\Big] + \dots \\
\label{chiMass}
\frac{\chi_a}{\Gamma} &= \frac{a_1}{b} \bigg\lbrace
\frac{GM}{b}\ms X_{1} 
+\Big(\frac{GM}{b}\Big)^2 \Big[\ms X_{2} + \delta \ms X_{2}^\delta\Big]
\nonumber\\
&\quad +\Big(\frac{GM}{b}\Big)^3\Big[\ms X_{3} + \delta \ms X_{3}^\delta +\nu \ms X_{3}^{\nu}\Big] \nonumber\\
&\quad
+\Big(\frac{GM}{b}\Big)^4\Big[\ms X_{4} + \delta \ms X_{4}^\delta +\nu \ms X_{4}^{\nu}+\nu\delta \ms X_{4}^{\nu\delta}\Big] \nonumber\\
&\quad + \Big(\frac{GM}{b}\Big)^5\Big[\ms X_{5} + \delta \ms X_{5}^\delta+\nu \ms X_{5}^{\nu}+\nu\delta \ms X_{5}^{\nu\delta}+\nu^2 \ms X_{5}^{\nu^2}\Big] \nonumber\\
&\quad + \dots \bigg\rbrace + 1 \leftrightarrow 2,
\end{align}
where the $\ms X_{\ms i}^\text{\dots}$ are functions only of energy/velocity.
Since $\nu$ and $\nu\delta$ are of order $q$ when expanded in the mass ratio, their coefficients can be recovered from 1SF results.

This mass-ratio dependence holds for the \emph{total} (local + nonlocal) scattering angle. However, by choosing the split between the local and nonlocal parts as we did in Sec.~\ref{sec:nonloc}, i.e., by choosing the arbitrary length scale $s$ to be the radial distance $r$, we get the same mass-ratio dependence for the \emph{local} part of the 5.5PN SO scattering angle. This is confirmed by the independent calculation of the nonlocal part of the scattering angle in Eq.~\eqref{chinonloc} below, which is linear in $\nu$. (In Ref.~\cite{Bini:2020wpo}, the authors introduced a `flexibility' factor in the relation between $s$ and $r$ to ensure that this mass-ratio dependence continues to hold at 5PN order for both the local and nonlocal contributions separately.)

Terms independent of $\nu$ in the scattering angle can be determined from the scattering angle of a spinning test particle in a Kerr background, which was calculated in Ref.~\cite{Bini:2017pee}. 
For a test body with spin $s$ in a Kerr background with spin $a$, the 5PM test-body scattering angle to all PN orders and to linear order in spins can be obtained by integrating Eq. (65) of Ref.~\cite{Bini:2017pee}, leading to
\begin{widetext}
\begin{align}
\chi_\text{test} &= \frac{GM}{b} \left[\frac{2v^2+2}{v^2}-\frac{4 (a+s)}{b v}\right]
+\pi\Big(\frac{GM}{b}\Big)^2 \left[ \frac{3 \left(v^2+4\right)}{4 v^2} -\frac{ \left(3 v^2+2\right) (4 a+3 s)}{2 b v^3}\right]  
\nonumber\\
&\quad
+ \Big(\frac{GM}{b}\Big)^3 \Bigg[\frac{2 \left(5 v^6+45 v^4+15 v^2-1\right)}{3 v^6} 
-\frac{4 \left(5 v^4+10 v^2+1\right) (3 a+2 s)}{b v^5}\Bigg] 
+\pi\Big(\frac{GM}{b}\Big)^4 \Bigg[\frac{105 \left(v^4+16 v^2+16\right)}{64 v^4} \nonumber\\
&\quad\qquad
-\frac{21  \left(5 v^4+20 v^2+8\right) (8 a+5 s)}{16 b v^5}\Bigg] 
+\Big(\frac{GM}{b}\Big)^5 \Bigg[\frac{2 \left(21 v^{10}+525 v^8+1050 v^6+210 v^4-15 v^2+1\right)}{5 v^{10}}\nonumber\\
&\quad\qquad
-\frac{4 \left(63 v^8+420 v^6+378 v^4+36 v^2-1\right) (5 a+3 s)}{3 b v^9}\Bigg] + \Order\left(G^6\right) + \Order(a^2,as,s^2).
\end{align}
Plugging this into Eq.~\eqref{chiMass} determines all the $\ms X_{\ms i}(v)$ and $\ms X_{\ms i}^\delta(v)$ functions. 
Hence, we can write the 5PM SO part of the local scattering angle, expanded to 5.5PN order, as follows:
\begin{align}
\label{chiAnz}
\frac{\chi_a^\text{loc}}{\Gamma} &= 
\frac{a_1}{b} \Bigg\lbrace \left(\frac{GM}{v^2b}\right) (-4 v) + \pi\left(\frac{GM}{v^2b}\right)^2 \left[\left(\frac{\delta }{2}-\frac{7}{2}\right) v + \left(\frac{3 \delta }{4}-\frac{21}{4}\right) v^3\right] \nonumber \\
&\quad
+\left(\frac{GM}{v^2b}\right)^3 \left[
(2 \delta -10 +  \ms X_{31}^\nu \nu) v + (20 \delta -100 +  \ms X_{33}^\nu \nu) v^3  +(10 \delta-50 +  \ms X_{35}^\nu \nu) v^5 + \ms X_{37}^\nu \nu v^7  + \ms X_{39}^\nu \nu v^9 
\right] \nonumber\\
&\quad 
+ \pi\left(\frac{GM}{v^2b}\right)^4 \bigg[
\left(\ms X_{41}^{\delta\nu} \delta \nu + \ms X_{41}^{\nu} \nu\right)v 
+\left(\frac{63}{4}\delta-\frac{273}{4} + \ms X_{43}^{\delta\nu} \delta \nu + \ms X_{43}^{\nu} \nu \right)v^3 
+ \left(\frac{315}{8}\delta-\frac{1365}{8} + \ms X_{45}^{\delta\nu} \delta  \nu  + \ms X_{45}^{\nu}\nu \right) v^5\nonumber \\
&\quad\qquad\qquad
+ \left(\frac{315 \delta }{32} -\frac{1365}{32} + \ms X_{47}^{\delta\nu} \delta \nu + \ms X_{47}^{\nu} \nu \right) v^7
+ \left(\ms X_{49}^{\delta\nu} \delta \nu + \ms X_{49}^{\nu} \nu\right) v^9
\bigg]\nonumber \\
&\quad
+\left(\frac{GM}{v^2b}\right)^5 \bigg[
\left(-\frac{4 \delta }{3}+\frac{16}{3}+ \ms X_{51}^{\delta\nu} \delta\nu+ \ms X_{51}^{\nu} \nu+ \ms X_{51}^{\nu^2} \nu^2\right) v
+\left(48 \delta -192 + \ms X_{53}^{\delta\nu} \delta  \nu +\ms X_{53}^{\nu} \nu +  \ms X_{53}^{\nu^2}\nu^2\right) v^3 \nonumber\\
&\quad\qquad\qquad
+ \left(504 \delta -2016 + \ms X_{55}^{\delta\nu} \delta  \nu +\ms X_{55}^{\nu} \nu +  \ms X_{55}^{\nu^2}\nu^2 \right) v^5
+ \left(560 \delta -2240 + \ms X_{57}^{\delta\nu} \delta  \nu +\ms X_{57}^{\nu} \nu +  \ms X_{57}^{\nu^2}\nu^2 \right) v^7 \nonumber\\
&\quad\qquad\qquad
+ \left(84\delta -336 + \ms X_{59}^{\delta\nu} \delta  \nu + \ms X_{59}^{\nu} \nu +  \ms X_{59}^{\nu^2} \nu^2\right)v^9
\bigg]\Bigg\rbrace + 1 \leftrightarrow 2,
\end{align}
\end{widetext}
where the $\ms X_{ij}^\nu$ and $\ms X_{ij}^{\delta\nu}$ coefficients are independent of the masses, and can be determined, as explained below, from 1SF results. The coefficient  $\ms X_{59}^{\nu^2}$ could be determined from future second-order self-force results.

\subsection{Relating the local Hamiltonian to the scattering angle}
The scattering angle can be calculated from the Hamiltonian by inverting the Hamiltonian and solving for $p_r(E,L,r)$, then evaluating the integral
\begin{equation}
\label{Htochi}
\chi= -2 \int_{r_0}^{\infty}\frac{\partial p_r(E,L,r)}{\partial L}dr - \pi\,,
\end{equation}
where $r_0$ is the turning point, obtained from the largest root of $p_r(E,L,r)=0$.
$E$ and $L$ represent the physical center-of-mass energy and \emph{canonical} angular momentum, respectively.

As noted above, we express the scattering angle in terms of the \emph{covariant} impact parameter $b$, but use the \emph{canonical} angular momentum $L$ in the Hamiltonian (corresponding to the Newton-Wigner spin-supplementary condition). The two are related via~\cite{Vines:2017hyw,Vines:2018gqi}
\begin{align}
\label{Lcov}
L&=L_\text{cov} + \Delta L, \nonumber\\
L_\text{cov} &= P_\text{c.m.} b = \frac{\mu}{\Gamma}\gamma vb, \nonumber\\
\Delta L &=M\frac{\Gamma-1}{2}\left[a_1+a_2-\frac{\delta}{\Gamma}(a_2-a_1)\right],
\end{align}
which can be used to replace $L$ with $b$ in the scattering angle.
We can also replace $E$ with $v$ using Eq.~\eqref{Egamma}.

Starting from the 4.5PN SO Hamiltonian, as given by Eq.~(5.6) of Ref.~\cite{Antonelli:2020ybz}, determines all the unknown coefficients in the scattering angle in Eq.~\eqref{chiAnz} up to that order.
Writing an ansatz for the local 5.5PN part in terms of unknown coefficients, such as
\begin{align}
g_S^\text{5.5PN,loc} = \frac{g_{04}}{r^4} + g_{23} \frac{p_r^2}{r^3} + g_{42} \frac{p_r^4}{r^2} + g_{61} \frac{p_r^6}{r} + g_{80} p_r^8, \nonumber\\
g_{S^*}^\text{5.5PN,loc} = \frac{g_{04}^*}{r^4} + g_{23}^* \frac{p_r^2}{r^3} + g_{42}^* \frac{p_r^4}{r^2} + g_{61}^* \frac{p_r^6}{r} + g_{80}^* p_r^8,
\end{align}
calculating the scattering angle, and matching to Eq.~\eqref{chiAnz} allows us to relate the 10 unknowns in that ansatz to the 6 unknowns in the scattering angle at that order. This leads to
\begin{widetext}
\begin{align}
\label{gyrosloc}
g_{S}^\text{5.5PN,loc} &= 2 \Bigg\lbrace\frac{1}{r^4} \Bigg[\nu  \left(\frac{3 \ms X_{59}^{\delta\nu}}{32}-2 \ms X_{49}^{\delta\nu}-\frac{35 \ms X_{39}^{\nu}}{16}+2 \ms X_{49}^{\nu}-\frac{3 \ms X_{59}^{\nu}}{32}+\frac{309077}{1152}-\frac{35449 \pi ^2}{3072}\right)+\nu^2 \left(\frac{235111}{2304}-\frac{3 \ms X_{59}^{\nu^2}}{32}-\frac{583 \pi ^2}{192}\right) \nonumber\\
&\qquad
-\frac{\nu ^4}{64}-\frac{413 \nu ^3}{512} \Bigg]
+\frac{p_r^2}{r^3} \left[\frac{3 \nu ^4}{8}-\frac{8259 \nu ^3}{128}+ \left(\frac{198133}{384}-\frac{1087 \pi ^2}{128}\right) \nu ^2+\nu  \left(2 \ms X_{49}^{\delta\nu}+\frac{35 \ms X_{39}^{\nu}}{8}-2 \ms X_{49}^{\nu}+\frac{1125}{16}\right)\right] \nonumber\\
&\quad
+\frac{p_r^4}{r^2} \left[-\frac{107 \nu ^4}{64}-\frac{73547 \nu ^3}{512}+\frac{31913 \nu ^2}{256}+\nu  \left(\frac{8597}{128}-\frac{35 \ms X_{39}^{\nu}}{24}\right)\right] \nonumber\\
&\quad
+\frac{p_r^6}{r} \left[\frac{1577 \nu ^4}{320}-\frac{11397 \nu ^3}{512}-\frac{2553 \nu ^2}{256}-\frac{893 \nu }{256}\right]
+ p_r^8 \left[\frac{189 \nu ^4}{64}+\frac{945 \nu ^3}{512}+\frac{99 \nu ^2}{256}-\frac{27 \nu }{128}\right]\Bigg\rbrace, \nonumber\\
g_{S^*}^\text{5.5PN,loc} &= \frac{3}{2}\Bigg\lbrace\frac{1}{r^4} \bigg[
-\frac{15 \nu ^4}{512}-\frac{111 \nu ^3}{128}+\nu  \left(2 \ms X_{49}^{\delta\nu}-\frac{3 \ms X_{59}^{\delta\nu}}{32}-\frac{35 \ms X_{39}^{\nu}}{16}+2 \ms X_{49}^{\nu}-\frac{3 \ms X_{59}^{\nu}}{32}+\frac{131519}{576}-\frac{90149 \pi ^2}{12288}\right)-\frac{1701}{512} \nonumber\\
&\qquad
+\nu ^2 \left(-\frac{3 \ms X_{59}^{\nu^2}}{32}-\frac{123 \pi ^2}{64}+\frac{29081}{512}\right)
\bigg]
+\frac{p_r^2}{r^3} \bigg[
\frac{171 \nu ^4}{256}-\frac{489 \nu ^3}{8}+\left(\frac{77201}{256}-\frac{123 \pi ^2}{16}\right) \nu ^2-\frac{27}{64}
 \nonumber\\
&\qquad
+\nu  \left(-2 \ms X_{49}^{\delta\nu}+\frac{35 \ms X_{39}^{\nu}}{8}-2 \ms X_{49}^{\nu}+\frac{86897}{768}-\frac{27697 \pi ^2}{2048}\right)\bigg]
+\frac{p_r^4}{r^2} \bigg[-\frac{1377 \nu ^4}{512}-\frac{13905 \nu ^3}{128}+\frac{12135 \nu ^2}{512}+\frac{2525}{512} \nonumber\\
&\qquad
+\nu  \left(\frac{10569}{256}-\frac{35 \ms X_{39}^{\nu}}{24}\right) \bigg]
+ \frac{p_r^6}{r}\left[\frac{16077 \nu ^4}{2560}-\frac{2391 \nu ^3}{640}-\frac{879 \nu ^2}{512}+\frac{77 \nu }{32}+\frac{3555}{512}\right] \nonumber\\
&\quad
+ p_r^8\left[\frac{945 \nu ^4}{512}+\frac{315 \nu ^3}{128}+\frac{1053 \nu ^2}{512}+\frac{189 \nu }{128}+\frac{693}{512}\right]\Bigg\rbrace,
\end{align}
\end{widetext}
where we switched to dimensionless variables.
We see that the 5 unknowns ($\ms X_{39}^\nu, \ms X_{49}^\nu,\ms X_{49}^{\delta\nu},\ms X_{59}^{\nu},\ms X_{59}^{\delta\nu}$) from the scattering angle only appear in the linear-in-$\nu$ coefficients of the gyro-gravitomagnetic factors up to order $p_r^4$, while the unknown $\ms X_{59}^{\nu^2}$ only appears in the quadratic-in-$\nu$ coefficients of the circular-orbit ($1/r^4$) part. All other coefficients have been determined, due to the structure of the PM-expanded scattering angle, and from lower-order and test-body results.

\subsection{Redshift and precession frequency}
To determine the linear-in-$\nu$ coefficients in the local Hamiltonian from 1SF results, we calculate the redshift and spin-precession invariants from the \emph{total} (local + nonlocal) Hamiltonian, since GSF calculations do not differentiate between the two, then match their small-mass-ratio expansion to 1SF expressions known in the literature.

An important step in this calculation is the first law of binary mechanics, which was derived for nonspinning particles in circular orbits in Ref.~\cite{LeTiec:2011ab}, generalized to spinning particles in circular orbits in Ref.~\cite{Blanchet:2012at}, to nonspinning particles in eccentric orbits in Refs.~\cite{Tiec:2015cxa,Blanchet:2017rcn}, and to spinning particles in eccentric orbits in Ref.~\cite{Antonelli:2020ybz}. It reads
\begin{equation}\label{1law}
\mr d E = \Omega_r \mr d I_r + \Omega_\phi \mr d L + \sum_{\mr i} ( z_{\mr i} \mr d m_{\mr i} + \Omega_{S_{\mr i}} \mr d S_{\mr i} ),
\end{equation}
where $\Omega_r$ and $\Omega_\phi$ are the radial and azimuthal frequencies, $I_r$ is the radial action, $z_{\mr i}$ is the redshift, and $ \Omega_{S_{\mr i}}$ is the spin-precession frequency.

The orbit-averaged redshift is a gauge-invariant quantity that can be calculated from the Hamiltonian using
\begin{equation}
z_1 = \left\langle \frac{\partial H}{\partial m_1} \right\rangle  = \frac{1}{T_r} \oint \frac{\partial H}{\partial m_1} dt, 
\end{equation}
where $T_r$ is the radial period.
The spin-precession frequency $\Omega_{S_1}$ and spin-precession invariant $\psi_1$ are given by
\begin{align}
\Omega_{S_1} &= \left\langle \frac{\partial H}{\partial S_1} \right\rangle = \frac{1}{T_r} \oint \frac{\partial H}{\partial S_1} dt, \nonumber\\
\psi_1 &\equiv \frac{ \Omega_{S_1}}{\Omega_\phi}.
\end{align}
In evaluating these integrals, we follow Refs.~\cite{Bini:2019lcd,Bini:2019lkm} in using the Keplerian parametrization for the radial variable
\begin{equation}
r = \frac{1}{u_p\left(1+e \cos \xi\right)},
\end{equation}
where $u_p$ is the inverse of the semi-latus rectum, $e$ is the eccentricity, and $\xi$ is the relativistic anomaly.
The radial and azimuthal periods are calculated from the Hamiltonian using
\begin{align}
T_r &\equiv \oint dt = 2 \int_0^\pi \left(\frac{\partial H}{\partial p_r}\right)^{-1} \frac{dr}{d\xi} d\xi, \\
T_\phi &\equiv \oint d\phi = 2 \int_0^\pi \frac{\partial H}{\partial L}\left(\frac{\partial H}{\partial p_r}\right)^{-1} \frac{dr}{d\xi} d\xi.
\end{align}

Performing the above steps yields the redshift and spin-precession invariants in terms of the gauge-dependent $u_p$ and $e$, i.e., $z_1(u_p, e)$ and $\psi_1(u_p, e)$.
We then express them in terms of the gauge-independent variables
\begin{equation}
\label{xiota}
x \equiv (M \Omega_\phi)^{2/3},   \quad  \iota \equiv \frac{3 x}{k},
\end{equation}
where $k\equiv T_\phi/(2\pi)-1$ is the fractional periastron advance.
The expressions we obtain for $z_1(x,\iota)$ and $\psi (x,\iota)$ agree up to 3.5PN order with those in Eq.~(50) of Ref.~\cite{Bini:2019lcd} and Eq.~(83) of Ref.~\cite{Bini:2019lkm}, respectively.

Note that the denominator of $\iota$ in Eq.~\eqref{xiota} is of order 1PN, which effectively scales down the PN ordering such that, to obtain the spin-precession invariant at fourth-subleading PN order, we need to include the 5PN nonspinning part of the Hamiltonian, which is given in Refs.~\cite{Bini:2019nra,Bini:2020wpo}.

\subsection{Comparison with self-force results}
Next, we expand the redshift $z_1(x, \iota)$ and spin-precession invariant $\psi_1(x, \iota)$ to first order in the mass ratio $q$, first order in the massive body's spin $a_2\equiv a$, and zeroth order in the spin of the smaller companion $a_1$. 
In doing so, we make use of another set of variables ($y,\lambda$), defined by
\begin{align}
\label{ylambda}
y &\equiv (m_2 \Omega_\phi)^{2/3} = \frac{x}{(1+q)^{2/3}}, \nonumber\\
\lambda &\equiv\frac{3y}{T_\phi/(2\pi)-1}  = \frac{\iota}{(1+q)^{2/3}} \,,
\end{align}
where the mass ratio $q=m_1/m_2$.

Schematically, those expansions have the following dependence on the scattering-angle unknowns:
\begin{align}
z_1(y,\lambda) &= \dots + q \left[\dots + a \left\{\ms X_{39}^\nu, \ms X_{49}^\nu-\ms X_{49}^{\delta\nu}, \ms X_{59}^\nu - \ms X_{59}^{\delta\nu}\right\}\right], \nonumber\\
\psi_1(y,\lambda) &= \dots + q \left\{\ms X_{39}^\nu, \ms X_{49}^\nu+\ms X_{49}^{\delta\nu}, \ms X_{59}^\nu + \ms X_{59}^{\delta\nu}\right\},
\end{align}
which can be seen from the structure of the scattering angle in Eq.~\eqref{chiAnz}.
In those expressions, the $\Order(a)$ part of the redshift depends on the unknown $\ms X_{39}^\nu$ and the \emph{difference} of the two pairs of unknowns $(\ms X_{49}^\nu,\ms X_{49}^{\delta\nu})$ and $(\ms X_{59}^\nu, \ms X_{59}^{\delta\nu})$, while the spin-precession invariant depends on $\ms X_{39}^\nu$ and the \emph{sum} of the two pairs of unknowns.
This means that solving for $\ms X_{39}^\nu$ requires 1SF result for \emph{either} $z_1$ or $\psi_1$, while solving for the other unknowns requires \emph{both}.

Hence, to solve for all five unknowns, we need at least three (or two) orders in eccentricity in the redshift, at first order in the Kerr spin, and two (or three) orders in eccentricity in the spin-precession invariant, at zeroth order in both spins.
Equivalently, instead of the spin-precession invariant, one could use the redshift at linear order in the spin of the smaller body $a_1$, but that is known from 1SF results for circular orbits only~\cite{Bini:2018zde}.
Incidentally, the available 1SF results are just enough to solve for the five unknowns, since the redshift is known to $\Order(e^4)$~\cite{Kavanagh:2016idg,Bini:2016dvs,Bini:2019lcd} and the spin-precession invariant to $\Order(e^2)$~\cite{Kavanagh:2017wot}.

The last unknown $\ms X_{59}^{\nu^2}$ in the 5.5PN scattering angle appears in both the redshift and spin-precession invariants at \emph{second} order in the mass ratio, thus requiring second-order self-force results for circular orbits. 

To compare $z_1(y,\lambda)$ and $\psi_1(y,\lambda)$ with GSF results, we write them in terms of the Kerr background values of the variables ($y,\lambda$) expressed in terms of $(u_p, e)$. The relations between the two sets of variables are explained in detail in Appendix~B of Ref.~\cite{Antonelli:2020ybz}, and we just need to append to Eqs.~(B16)-(B20) there the following PN terms
\begin{align}
y(u_p,e)&=y_{0}(u_p,e)+ a\, y_{a}(u_p,e)+\mathcal{O}(a^2), \\
\lambda(u_p,e)&=\lambda_{0}(u_p,e)+ a\, \lambda_{a}(u_p,e)+\mathcal{O}(a^2),\nonumber\\
y_a(u_p,e)&= \dots + \left(\frac{4829 e^4}{12}-4984 e^2\right) u_p^{13/2}, \nonumber\\
\lambda_a(u_p,e)&= \dots + \left(\frac{4671}{8}+\frac{13959 e^2}{8}-\frac{19657 e^4}{12}\right) u_p^{11/2}.\nonumber
\end{align}

We obtain the following 1SF part of the inverse redshift $U_1 \equiv 1/z_1$ and spin-precession invariant $\psi_1$
\begin{align}
U_1&=  U^{(0)}_{1a^0} +a\, U^{(0)}_{1a}+q\left(U^\text{1SF}_{1a^0} +a\, U^\text{1SF}_{1a}\right)+\mathcal{O}(q^2,a^2), \nonumber\\
\psi_1&=  \psi^{(0)}_{1a^0} +q \psi^\text{1SF}_{1a^0} +\mathcal{O}(q^2,a)\,,
\end{align}
\begin{widetext}
\begin{align}
\label{U1SF}
U^\text{1SF}_{1a}&=\left(3-\frac{7 e^2}{2}-\frac{e^4}{8}\right) u_p^{5/2}+\left(18-4
e^2-\frac{117 e^4}{4}\right) u_p^{7/2} +\left(87+\frac{287 e^2}{2}-\frac{6277 e^4}{16}\right)u_p^{9/2}\nonumber\\
&\quad +\left[\frac{3890}{9}-\frac{241 \pi ^2}{96}+\left(\frac{5876}{3}-\frac{569 \pi ^2}{64}\right) e^2+\left(\frac{2025 \pi ^2}{128}-3547\right) e^4\right]u_p^{11/2} \nonumber\\
&\quad
+ \bigg[
8 \ms X_{49}^{\delta\nu}-\frac{3 \ms X_{59}^{\delta\nu}}{8}+\frac{35 \ms X_{39}^{\nu}}{4}-8 \ms X_{49}^{\nu}+\frac{3 \ms X_{59}^{\nu}}{8}+\frac{17917 \pi ^2}{768}+\frac{2027413}{2880}+\frac{2336 \gamma_E }{15}+\frac{928 \ln 2}{3} + \frac{1168 \ln u_p}{15} \nonumber\\
&\quad\qquad
+ e^2 \bigg(
\frac{4832 \ln u_p}{15}+24 \ms X_{49}^{\delta\nu}-\frac{21 \ms X_{59}^{\delta\nu}}{16}+\frac{175 \ms X_{39}^{\nu}}{8}-24 \ms X_{49}^{\nu}+\frac{21 \ms X_{59}^{\nu}}{16}+\frac{182411 \pi ^2}{1536}+\frac{31389241}{2880}+\frac{9664 \gamma_E }{15} \nonumber\\
&\qquad\qquad
-1728 \ln 2+2916 \ln 3
\bigg) 
+ e^4 \bigg(-\frac{1248 \ln u_p}{5}-42 \ms X_{49}^{\delta\nu}+\frac{63 \ms X_{59}^{\delta\nu}}{32}-\frac{175 \ms X_{39}^{\nu}}{4}+42 \ms X_{49}^{\nu}-\frac{63 \ms X_{59}^{\nu}}{32}-\frac{2496 \gamma_E }{5} \nonumber\\
&\qquad\qquad
-\frac{200393 \pi ^2}{1024}-\frac{137249131}{7680}+\frac{782912 \ln 2}{15}-\frac{328779 \ln 3}{10}\bigg)
\bigg]u_p^{13/2}
\,, \\
\label{psi1SF}
\psi_{1\, a^0}^\text{1SF} &=-u_p+\left(\frac{9}{4}+e^2\right) u_p^2+
\left[\frac{739}{16}-\frac{123 \pi ^2}{64}+\left(\frac{341}{16}-\frac{123 \pi ^2}{256}\right) e^2\right]u_p^3\nonumber\\
&\quad+\bigg[\frac{628 \ln u_p}{15}+\frac{31697 \pi ^2}{6144}-\frac{587831}{2880}+\frac{1256 \gamma_E }{15} + \frac{296}{15} \ln 2+\frac{729 \ln 3}{5}\nonumber\\
&\quad\qquad
+e^2
\bigg(\frac{268 \ln u_p}{5}-\frac{164123}{480}-\frac{23729 \pi ^2}{4096}+\frac{536 \gamma_E }{5}+\frac{11720 \ln 2}{3}-\frac{10206 \ln 3}{5}\bigg)\bigg]u_p^4 \nonumber\\
&\quad
+ \bigg[
4 \ms X_{49}^{\delta\nu}-\frac{3 \ms X_{59}^{\delta\nu}}{16}-\frac{35 \ms X_{39}^{\nu}}{8}+4 \ms X_{49}^{\nu}-\frac{3 \ms X_{59}^{\nu}}{16}+\frac{6793111 \pi ^2}{24576}-\frac{22306 \gamma_E }{35}-\frac{115984853}{57600}+\frac{22058 \ln 2}{105}-\frac{31347 \ln 3}{28} \nonumber\\
&\qquad\quad
-\frac{11153}{35} \ln u_p 
+ e^2 \bigg(\frac{4248047}{4800}+18 \ms X_{49}^{\delta\nu}-\frac{15 \ms X_{59}^{\delta\nu}}{16}-\frac{35 \ms X_{39}^{\nu}}{2}+18 \ms X_{49}^{\nu}-\frac{15 \ms X_{59}^{\nu}}{16}+\frac{4895607 \pi ^2}{16384}-\frac{22682 \gamma_E }{15} \nonumber\\
&\qquad\quad
+\frac{4430133 \ln 3}{320}+\frac{9765625 \ln 5}{1344}-\frac{4836254 \ln 2}{105}-\frac{11341 \ln u_p}{15}\bigg)
\bigg]u_p^5
\,.
\end{align}
\end{widetext}

These results can be directly compared with those derived in GSF literature. In particular, for the redshift, we match to Eq.~(4.1) of Ref.~\cite{Kavanagh:2016idg}, Eq.~(23) of Ref.~\cite{Bini:2016dvs}, and Eq.~(20) of Ref.~\cite{Bini:2019lcd}, while for the precession frequency, we match to Eq.~(3.33) of Ref.~\cite{Kavanagh:2017wot}\footnote{
Note that the $\Order(e^2 u_p^5)$ term in Eq.~(3.33) of Ref.~\cite{Kavanagh:2017wot} has a typo, but the correct expression is provided in the Black Hole Perturbation Toolkit~\cite{BHPToolkit}.
}.
The matching to 1SF results leads to the following solution for the unknown coefficients in the scattering angle:
\begin{align}
\label{Xsol}
\ms X_{39}^{\nu} &= \frac{26571}{1120}, \nonumber\\
\ms X_{49}^{\delta\nu} &= \frac{533669}{4800}-\frac{97585 \pi ^2}{8192}, \nonumber\\
\ms X_{49}^{\nu} &= -\frac{403129}{4800}+\frac{80823 \pi ^2}{8192}, \nonumber\\
\ms X_{59}^{\delta\nu} &= \frac{285673}{240}-\frac{2477 \pi ^2}{16}, \nonumber\\
\ms X_{59}^{\nu} &= \frac{402799}{270}-\frac{4135 \pi ^2}{144}.
\end{align}
Note that all the logarithms and Euler constants, which are purely nonlocal, cancel between GSF results and those in Eqs.~\eqref{U1SF} and \eqref{psi1SF}, thus providing a good check for our calculations.

Another check would be possible once 1SF results are computed at higher orders in eccentricity, since one could directly compare them to our results for the redshift and spin-precession invariants that are provided in the  Supplemental Material~\cite{ancprd} expanded to $\Order(e^8)$.

\subsection{Local scattering angle and Hamiltonian}
Inserting the solution from Eq.~\eqref{Xsol} into the scattering angle in Eq.~\eqref{chiAnz}, yields
\begin{widetext}
\begin{align}
\label{chiLoc}
\frac{\chi_a^\text{loc}}{\Gamma} &= 
\frac{a_1}{b} \Bigg\lbrace \left(\frac{GM}{v^2b}\right) (-4 v) + \pi\left(\frac{GM}{v^2b}\right)^2 \left[\left(\frac{\delta }{2}-\frac{7}{2}\right) v + \left(\frac{3 \delta }{4}-\frac{21}{4}\right) v^3\right] \nonumber \\
&\quad
+\left(\frac{GM}{v^2b}\right)^3 \left[
(2 \delta -10) v + (20 \delta -100 +  10 \nu) v^3  +\left(10 \delta-50 + \frac{77}{2} \nu\right) v^5 + \frac{177}{4} \nu v^7  + \frac{26571}{1120} \nu v^9 
\right] \nonumber\\
&\quad 
+ \pi\left(\frac{GM}{v^2b}\right)^4 \bigg[
\left(\frac{63}{4}\delta-\frac{273}{4} - \frac{3}{4} \delta \nu + \frac{39}{4} \nu \right)v^3 
+ \left(\frac{315}{8}\delta-\frac{1365}{8} - \frac{45}{8} \delta  \nu  + \frac{777}{8} \nu \right) v^5\nonumber \\
&\qquad\qquad
+ \left(\frac{315 \delta }{32} -\frac{1365}{32} + \left(-\frac{257}{96}-\frac{251 \pi ^2}{256}\right) \delta \nu + \left(\frac{23717}{96}-\frac{733 \pi ^2}{256}\right) \nu \right) v^7 \nonumber\\
&\qquad\qquad
+ \left(\left(\frac{533669}{4800}-\frac{97585 \pi ^2}{8192}\right) \delta \nu + \left(\frac{80823 \pi ^2}{8192}-\frac{403129}{4800}\right) \nu\right) v^9
\bigg]\nonumber \\
&\quad
+\left(\frac{GM}{v^2b}\right)^5 \bigg[
\left(-\frac{4 \delta }{3}+\frac{16}{3}\right) v
+\left(48 \delta -192 - 4 \delta  \nu + 32 \nu\right) v^3
+ \left(504 \delta -2016 - 109 \delta  \nu + 1032 \nu - 16 \nu^2 \right) v^5 \nonumber\\
&\qquad\qquad
+ \left(560 \delta -2240 + \left(-\frac{21995}{54}-\frac{80 \pi ^2}{9}\right) \delta  \nu + \left(\frac{150220}{27}-\frac{2755 \pi ^2}{36}\right) \nu - 168 \nu^2 \right) v^7 \nonumber\\
&\qquad\qquad
+ \left(84\delta -336 + \left(\frac{285673}{240}-\frac{2477 \pi ^2}{16}\right) \delta  \nu + \left(\frac{402799}{270}-\frac{4135 \pi ^2}{144}\right) \nu +  \ms X_{59}^{\nu^2} \nu^2\right)v^9
\bigg]\Bigg\rbrace + 1 \leftrightarrow 2.
\end{align}

For the gyro-gravitomagnetic factors, which are one of the main results of this paper, substituting the solution~\eqref{Xsol} in Eq.~\eqref{gyrosloc} yields the following local part:
\begin{align}
\label{gSLoc}
g_S^\text{5.5PN,loc} &= 2\Bigg\lbrace\left[-\frac{\nu ^4}{64}-\frac{413 \nu ^3}{512}+\nu ^2 \left(-\frac{3 \ms X_{59}^{\nu^2}}{32}-\frac{583 \pi ^2}{192}+\frac{235111}{2304}\right)+\left(\frac{62041 \pi ^2}{3072}-\frac{11646877}{57600}\right) \nu \right]\frac{1}{r^4} \nonumber\\
&\quad 
+ \left[\frac{3 \nu ^4}{8}-\frac{8259 \nu ^3}{128}+\left(\frac{198133}{384}-\frac{1087 \pi ^2}{128}\right) \nu ^2+\left(\frac{3612403}{6400}-\frac{22301 \pi ^2}{512}\right) \nu \right] \frac{p_r^2}{r^3} \nonumber\\
&\quad
+ \left[-\frac{107 \nu ^4}{64}-\frac{73547 \nu ^3}{512}+\frac{31913 \nu ^2}{256}+\frac{8337 \nu }{256}\right] \frac{p_r^4}{r^2}
+ \left[\frac{1577 \nu ^4}{320}-\frac{11397 \nu ^3}{512}-\frac{2553 \nu ^2}{256}-\frac{893 \nu }{256}\right] \frac{p_r^6}{r} \nonumber\\
&\quad
+ \left[\frac{189 \nu ^4}{64}+\frac{945 \nu ^3}{512}+\frac{99 \nu ^2}{256}-\frac{27 \nu }{128}\right]  p_r^8\Bigg\rbrace, \\
\label{gSstrLoc}
g_{S^*}^\text{5.5PN,loc} &= \frac{3}{2} \Bigg\lbrace
\left[-\frac{5 \nu ^4}{128}-\frac{37 \nu ^3}{32}+\nu ^2 \left(-\frac{\ms X_{59}^{\nu^2}}{8}-\frac{41 \pi ^2}{16}+\frac{29081}{384}\right)+\left(\frac{23663 \pi ^2}{3072}-\frac{55}{2}\right) \nu -\frac{567}{128}\right]\frac{1}{r^4} \nonumber\\
&\quad
+  \left[\frac{57 \nu ^4}{64}-\frac{163 \nu ^3}{2}+\left(\frac{77201}{192}-\frac{41 \pi ^2}{4}\right) \nu ^2+\left(\frac{34677}{160}-\frac{4829 \pi ^2}{384}\right) \nu -\frac{9}{16}\right] \frac{p_r^2}{r^3} \nonumber\\
&\quad
+  \left[-\frac{459 \nu ^4}{128}-\frac{4635 \nu ^3}{32}+\frac{4045 \nu ^2}{128}+\frac{107 \nu }{12}+\frac{2525}{384}\right] \frac{p_r^4}{r^2}
+ \left[\frac{5359 \nu ^4}{640}-\frac{797 \nu ^3}{160}-\frac{293 \nu ^2}{128}+\frac{77 \nu }{24}+\frac{1185}{128}\right] \frac{p_r^6}{r} \nonumber\\
&\quad
+  \left[\frac{315 \nu ^4}{128}+\frac{105 \nu ^3}{32}+\frac{351 \nu ^2}{128}+\frac{63 \nu }{32}+\frac{231}{128}\right]p_r^8\Bigg\rbrace.
\end{align}
\end{widetext}

\subsection{Comparison with numerical relativity}
\label{sec:Eb}
To quantify the effect of the 5.5PN SO part on the dynamics, and that of the remaining unknown coefficient $\ms X_{59}^{\nu^2}$, we compare the binding energy calculated from the EOB Hamiltonian to NR.
The binding energy provides a good diagnostic for the conservative dynamics of the binary system~\cite{Barausse:2011dq,Damour:2011fu,Nagar:2015xqa}, and can be calculated from accurate NR simulations by subtracting the radiated energy $E_\text{rad}$ from the ADM energy $E_\text{ADM}$ at the beginning of the simulation~\cite{Ruiz:2007yx}, i.e.,
\begin{equation}
\bar{E}_\text{NR} = E_\text{ADM} - E_\text{rad} - M.
\end{equation}

To isolate the SO contribution $\bar{E}^\text{SO}$ to the binding energy, we combine configurations with different spin orientations (parallel or anti-parallel to the orbital angular momentum), as explained in Refs.~\cite{Dietrich:2016lyp,Ossokine:2017dge}. One possibility is to use
\begin{equation}
\label{EbSO}
\bar{E}^\text{SO}(\nu,\chi,\chi) \simeq \frac{1}{2} \left[ \bar{E}(\nu,\chi,\chi) - \bar{E}(\nu,-\chi,-\chi)\right],
\end{equation}
where $\chi$ here is the magnitude of the dimensionless spin.
This relation subtracts the nonspinning and spin-spin parts, with corrections remaining at order $\chi^3$, which provides a good approximation since the spin-cubed contribution to the binding energy is about an order of magnitude smaller than the SO contribution, as was shown in Ref.~\cite{Ossokine:2017dge}.

We calculate the binding energy for circular orbits from the EOB Hamiltonian using $\bar{E}_\text{EOB} = H_\text{EOB} - M$ while neglecting radiation reaction effects, which implies that $\bar{E}_\text{EOB}$ is not expected to agree well with $\bar{E}_\text{NR}$ near the end of the inspiral. 
We set $p_r=0$ in the Hamiltonian and numerically solve $\dot{p}_r=0=-\partial H/\partial r$ for the angular momentum $L$ at different orbital separations.
Then, we plot $\bar{E}$ versus the dimensionless parameter
\begin{equation}
v_\Omega \equiv (M\Omega)^{1/3},
\end{equation}
where the orbital frequency $\Omega = \partial H / \partial L$.
Finally, we compare the EOB binding energy to NR data for the binding energy that were extracted in Ref.~\cite{Ossokine:2017dge} from the Simulating eXtreme Spacetimes (SXS) catalog~\cite{SXS,Boyle:2019kee}. In particular, we use the simulations with SXS ID 0228 and 0215 for $q=1$, 0291 and 0264 for $q=1/3$, all with spin magnitudes $\chi=0.6$ aligned and antialigned. The numerical error in these simulations is significantly smaller than the SO contribution to the binding energy.

\begin{figure}
\includegraphics{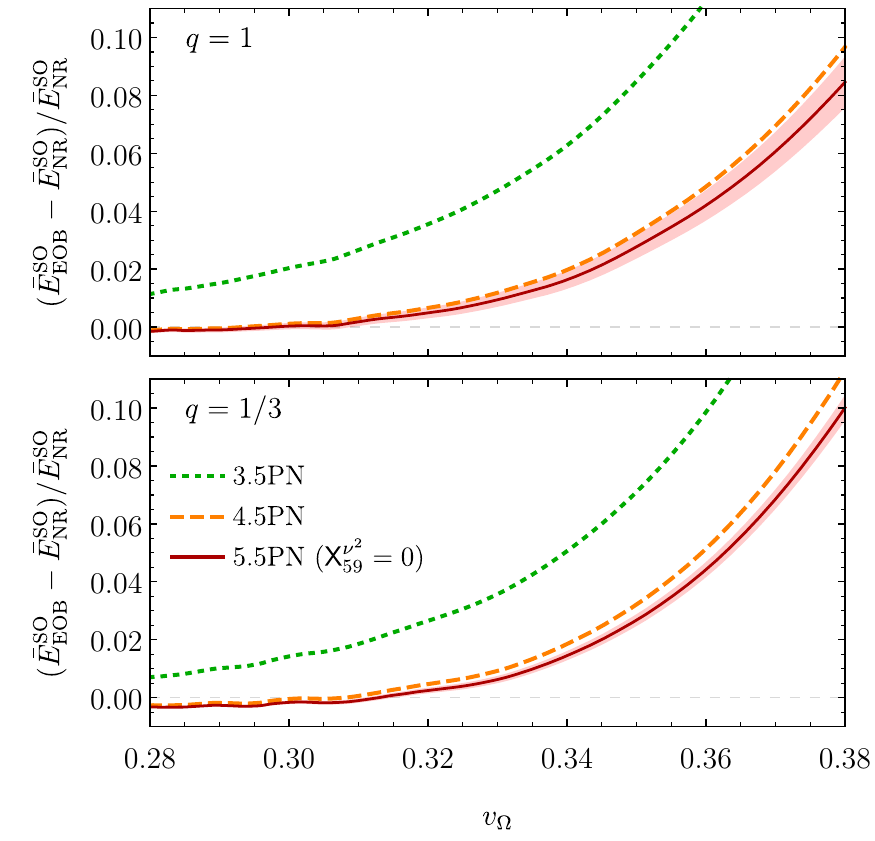}
\caption{The relative difference in the SO contribution to the binding energy between EOB and NR, plotted versus the frequency parameter $v_\Omega$. The 5.5PN curve corresponds to $\ms X_{59}^{\nu^2} = 0$, while the upper and lower edges of the shaded region around it correspond to  $\ms X_{59}^{\nu^2} = -500$ and $\ms X_{59}^{\nu^2} = 500$, respectively.}
\label{fig:Eb}
\end{figure}

In Fig.~\ref{fig:Eb}, we plot the relative difference in the SO contribution $\bar{E}^\text{SO}$ between EOB and NR for two mass ratios, $q=1$ and $q=1/3$, as a function of $v_\Omega$ up to $v_\Omega = 0.38$, which corresponds to about an orbit before merger.
We see that the inclusion of the 5.5PN SO part (with the remaining unknown $\ms X_{59}^{\nu^2} = 0$) provides an improvement over 4.5PN, but the difference is smaller than that between 3.5PN and 4.5PN.
In addition, since the remaining unknown $\ms X_{59}^{\nu^2}$ is expected to be about $\Order(10^2)$, based on the other coefficients in the scattering angle, we plotted the energy for $\ms X_{59}^{\nu^2}=500$ and $\ms X_{59}^{\nu^2}=-500$, demonstrating that the effect of that unknown is less than the difference between 4.5PN and 5.5PN, with decreasing effect for small mass ratios.

\section{Nonlocal 5.5PN SO scattering angle}
\label{sec:nonlocscatter}
The local part of the Hamiltonian and scattering angle calculated in the previous section is valid for both bound and unbound orbits. However, the nonlocal part of the Hamiltonian from Sec.~\ref{sec:nonloc} is only valid for bound orbits since it was calculated in a small-eccentricity expansion.
In this section, we complement these results by calculating the nonlocal part for unbound orbits in a large-eccentricity (or large-angular-momentum) expansion.

The nonlocal part of the 4PN scattering angle was first computed in Ref.~\cite{Bini:2017wfr}, in both the time and frequency domains, at leading order in the large-eccentricity expansion.
This was extended in Ref.~\cite{Bini:2020wpo} at 5PN at leading order in eccentricity, and in Ref.~\cite{Bini:2020hmy} at 6PN to next-to-next-to-leading order in eccentricity.
In addition, Refs.~\cite{Bini:2020uiq,Bini:2020rzn} recovered analytical expressions for the nonlocal scattering angle by using high-precision arithmetic methods.

It was shown in Ref.~\cite{Bini:2017wfr} that the nonlocal contribution to the scattering angle is given by
\begin{align}
\chi_\text{nonloc} &= \frac{1}{\nu} \frac{\partial}{\partial L} W_\text{nonloc}(E,L), \\
W_\text{nonloc} &= \int dt\, \delta H_\text{nonloc},
\end{align}
with $\delta H_\text{nonloc}$ given by Eq.~\eqref{nonlocH}, leading to
\begin{align}
W_\text{nonloc} &= W^\text{flux split} + W^\text{flux}, \\
W^\text{flux split} &= - \frac{G M}{c^3} \int dt\, \text{Pf}_{2s/c} \int \frac{d\tau}{|\tau|} \mathcal{F}_\text{LO}^\text{split}(t,t+\tau), \label{Wsplit} \\
W^\text{flux} &= \frac{2GM}{c^3} \int dt\, \mathcal{F}_\text{LO}(t,t) \ln\left(\frac{r}{s}\right). \label{Wflux}
\end{align}

To evaluate the integral in the large-eccentricity limit, we follow the steps used in Refs.~\cite{Bini:2017wfr,Bini:2020wpo}. 
We use the quasi-Keplerian parametrization for hyperbolic motion~\cite{damour1985general,Cho:2018upo}
\begin{align}
r &= \bar{a}_r (e_r \cosh \bar{u} - 1), \\
\bar{n} t &= e_t \sinh \bar{u} - \bar{u}, \label{KepEqhyp}\\
\phi &= 2 K \arctan\left[\sqrt{\frac{e_\phi+1}{e_\phi-1}} \tanh \frac{\bar{u}}{2}\right], \label{qKphihyp}
\end{align}
which is the analytic continuation of the parametrization for elliptic orbits in Eqs.~\eqref{qKr}--\eqref{qKphi}. In Appendix.~\ref{app:qKephyp}, we summarize the relations for these quantities in terms of the energy and angular momentum.

We begin by expressing the variables $(r,\dot{\phi},\dot{r})$, which enter the multipole moments, in terms of $(\phi,L,e_t)$, such that
\begin{align}
r &= \frac{L^2}{1 + e_t \cos\phi} +\frac{2 \delta  \chi _A+(2-\nu ) \chi _S}{L \left(e_t \cos \phi +1\right)^2} \nonumber\\
&\qquad 
\times \left(2 \phi  e_t \sin \phi+4 e_t \cos\phi +e_t^2+3\right), \nonumber\\
\dot{\phi} &= \frac{\left(e_t \cos \phi +1\right)^2}{L^3} +\frac{\left(e_t \cos \phi +1\right) \left(2 \delta  \chi_A+(2-\nu ) \chi _S\right)}{2 L^6} \nonumber\\
&\quad
\times  \left(-8 \phi  e_t \sin \phi +e_t^2 \cos (2 \phi )-12 e_t \cos \phi-3 e_t^2-10\right), \nonumber\\
\dot{r} &= \frac{e_t \sin \phi}{L} +\frac{e_t \left(2 \delta  \chi _A+(2-\nu) \chi _S\right)}{2 L^4} \nonumber\\
&\qquad
\times \left(e_t \sin (2 \phi )-2 \sin \phi +4 \phi  \cos \phi \right).
\end{align}

We then use these relations to obtain an exact expression for the flux-split function $\mathcal{F}_\text{LO}^\text{split}(\phi,\phi')$, with  no eccentricity expansion, which takes the form
\begin{align}
\label{Fsplitphi}
\mathcal{F}_\text{LO}^\text{split}(\phi,\phi') &= \frac{4\nu^2}{15L^{10}} (1+e_t\cos\phi)^2 (1+e_t\cos\phi')^2 \nonumber\\
&\qquad 
\times \left(F_0 +F_1 e_t + F_2 e_t^2\right) \nonumber\\
&\quad + \frac{\nu^2}{L^{13}} (1+e_t\cos\phi) (1+e_t\cos\phi') \nonumber\\
&\qquad
\times \left(F_0^s +F_1^s e_t + \dots + F_6^s e_t^6\right),
\end{align}
where the functions $F_{\ms i}(\phi,\phi')$ are given by Eq.~(92) of Ref.~\cite{Bini:2017xzy}, but the functions $F_{\ms i}^s(\phi,\phi')$ in the SO part are too lengthy to write here. 
Instead, we expand $\mathcal{F}_\text{LO}^\text{split}(\phi,\phi')$ to leading order in a large-eccentricity expansion (in powers of $1/e_t$). 
To do that, we define the rescaled mean motion $\tilde{n} \equiv \bar{n} / e_t$, write the Kepler Eq.~\eqref{KepEqhyp} as
\begin{equation}
\tilde{n}t = \sinh \bar{u} - \frac{\bar{u}}{e_t},
\end{equation}
and solve for $\bar{u}$ in a  $1/e_t$ expansion
\begin{equation}
\bar{u} = \sinh ^{-1}\left(t \tilde{n}\right) + \frac{\sinh ^{-1}\left(t \tilde{n}\right)}{e_t \sqrt{1+t^2 \tilde{n}^2}} + \Order(e_t^{-2}).
\end{equation}

Substituting in Eq.~\eqref{qKphihyp} and expanding yields
\begin{align}
\phi(t) &= \tan ^{-1}\left(t \tilde{n}\right)+ \Order\left(e_t^{-1}\right) \nonumber\\
&\quad - \frac{t \tilde{n} e_t \left[2 \delta  \chi _A+(2-\nu) \chi _S\right]}{L^3 \sqrt{t^2 \tilde{n}^2+1}} + \Order\left(e_t^0\right).
\end{align}
Defining $\tilde{t} \equiv \tilde{n} t$ and $\tilde{\tau} \equiv \tilde{n} \tau$, then substituting in Eq.~\eqref{Fsplitphi} and expanding yields
\begin{widetext}
\begin{align}
\mathcal{F}^\text{split}(t,t+\tau) &= \frac{4 \nu^2 e_t^6 }{15 L^{10}} \frac{\ms f_6(\tilde{t},\tilde{\tau})}{\left(\tilde{t}^2+1\right)^{5/2} \left(2 \tilde{\tau } \tilde{t}+\tilde{t}^2+\tilde{\tau }^2+1\right)^{5/2}} + \Order(e_t^5) \nonumber\\
&\quad
+ \frac{4 \nu^2 e_t^8 }{15 L^{13} } \frac{\chi_S \ms f_8^S(\tilde{t},\tilde{\tau}) + \delta \chi_A \ms f_8^A(\tilde{t},\tilde{\tau})}{\left(\tilde{t}^2+1\right)^{7/2} \left(2 \tilde{\tau } \tilde{t}+\tilde{t}^2+\tilde{\tau }^2+1\right)^{7/2}}+ \Order(e_t^7),
\end{align}
with
\begin{align}
\ms f_6(\tilde{t},\tilde{\tau}) &=2 \tilde{t}^6+ 6 \tilde{\tau } \tilde{t}^5+\left(6 \tilde{\tau }^2+28\right) \tilde{t}^4+2 \tilde{\tau } \left(\tilde{\tau }^2+28\right) \tilde{t}^3+\left(39 \tilde{\tau }^2+50\right) \tilde{t}^2+\tilde{\tau } \left(11 \tilde{\tau }^2+50\right) \tilde{t}-12 \left(\tilde{\tau }^2-2\right),
\nonumber\\
\ms f_8^S(\tilde{t},\tilde{\tau}) &= 4 (9 \nu -5) \tilde{t}^8 + 16 (9 \nu -5) \tilde{\tau } \tilde{t}^7+2 \tilde{t}^6 \left[18 \nu  \left(6 \tilde{\tau }^2+5\right)-63 \tilde{\tau }^2-67\right]+2 \tilde{\tau } \tilde{t}^5 \left[18 \nu  \left(4 \tilde{\tau }^2+15\right)-49 \tilde{\tau }^2-201\right] \nonumber\\
&\quad
+\tilde{t}^4 \left[18 \nu  \left(2 \tilde{\tau }^4+35 \tilde{\tau }^2+18\right)-2 \left(19 \tilde{\tau }^4+206 \tilde{\tau }^2+141\right)\right]-2 \tilde{\tau } \tilde{t}^3 \left[-36 \nu  \left(5 \tilde{\tau }^2+9\right)+3 \tilde{\tau }^4+77 \tilde{\tau }^2+282\right] \nonumber\\
&\quad
+\tilde{t}^2 \left[3 \nu  \left(35 \tilde{\tau }^4+130 \tilde{\tau }^2+84\right)-2 \left(8 \tilde{\tau }^4+169 \tilde{\tau }^2+121\right)\right]
+\tilde{\tau } \tilde{t} \left[3 \nu  \left(5 \tilde{\tau }^4+22 \tilde{\tau }^2+84\right)-2 \left(3 \tilde{\tau }^4+28 \tilde{\tau }^2+121\right)\right] \nonumber\\
&\quad
+22 \tilde{\tau }^4-52 \tilde{\tau }^2-74 -12 \nu  \left(3 \tilde{\tau }^4+2 \tilde{\tau }^2-6\right),
\nonumber\\
\ms f_8^A(\tilde{t},\tilde{\tau}) &= -2  \left(\tilde{t}^2+1\right) \Big[40 \tilde{\tau } \tilde{t}^5+\left(63 \tilde{\tau }^2+57\right) \tilde{t}^4+7 \tilde{\tau } \left(7 \tilde{\tau }^2+23\right) \tilde{t}^3+\left(19 \tilde{\tau }^4+143 \tilde{\tau }^2+84\right) \tilde{t}^2+\tilde{\tau } \left(3 \tilde{\tau }^4+28 \tilde{\tau }^2+121\right) \tilde{t} \nonumber\\
&\quad\qquad
+10 \tilde{t}^6-11 \tilde{\tau }^4+26 \tilde{\tau }^2+37 \Big].
\end{align}

Then, we evaluate the \emph{partie finie} integral in Eq.~\eqref{Wsplit}, after writing it in terms of $\tilde{T} \equiv 2\tilde{n}s/c$, to obtain
\begin{align}
&-\text{Pf}_{\tilde{T}} \int \frac{d\tilde{\tau}}{|\tilde{\tau}|} \mathcal{F}_\text{LO}^\text{split}(t,t+\tau)
= \frac{8 \nu^2 e_t^6 }{15 L^{10} \left(\tilde{t}^2+1\right)^3} \left[36 + 7 \tilde{t}^2 + 2 \left(\tilde{t}^2+12\right) \ln \left(\frac{\tilde{T}}{2 \tilde{t}^2+2}\right)\right] \nonumber\\
&\qquad\qquad
+\frac{16 \nu^2 e_t^8}{15 L^{13} \left(\tilde{t}^2+1\right)^4} \Bigg\lbrace 
\chi _S \left[(3 \nu -5) \tilde{t}^4+(36 \nu +5) \tilde{t}^2-\left(2 (5-9 \nu ) \tilde{t}^2-36 \nu +37\right) \ln \left(\frac{\tilde{T}}{2\tilde{t}^2+2}\right)+60 \nu -53\right]
 \nonumber\\
&\qquad\qquad\qquad
-\delta  \chi _A \left[\left(10 \tilde{t}^2+37\right) \ln \left(\frac{\tilde{T}}{2\tilde{t}^2+2}\right)+5 \tilde{t}^4-5 \tilde{t}^2+53\right]
\Bigg\rbrace.
\end{align}
Integrating over $t$, we obtain the flux-split potential
\begin{align}
W^\text{flux split} &= \frac{2\pi \nu^2}{15 e_t^3 a_r^{7/2}} \left[100 + 37 \ln\left(\frac{s}{4 e_t a_r^{3/2}}\right)\right]
+\frac{\pi \nu^2}{30 e_t^4 a_r^5} \Bigg\lbrace
31\delta \chi_A \left[-137 - 46 \ln \left(\frac{s}{4 a_r^{3/2} e_t}\right) \right] \nonumber\\
&\qquad
+ \chi_S \left[
2774 \nu -4247 - 2(713-457 \nu ) \ln \left(\frac{s}{4 a_r^{3/2} e_t}\right)
\right]
\Bigg\rbrace.
\end{align}
The second contribution $W^\text{flux}$ in Eq.~\eqref{Wflux} can be easily integrated to yield
\begin{align}
W^\text{flux} &= \frac{2\pi \nu^2}{15 e_t^3 a_r^{7/2}} \left[-\frac{85}{4} - 37 \ln\left(\frac{s}{2 a_r e_t}\right)\right] 
+\frac{\pi \nu^2}{60 e_t^4 a_r^5} \bigg\lbrace
\delta \chi_A \left[2255 + 2852\ln\left( \frac{s}{2 a_r e_t}\right) \right] \nonumber\\
&\qquad
+ \chi_S \left[
2255-1365 \nu + (2852-1828 \nu ) \ln \left(\frac{s}{2 a_r e_t}\right)
\right]
\bigg\rbrace,
\end{align}
\end{widetext}
where we used
\begin{equation}
r(t) = \frac{L^2 \sqrt{\tilde{t}^2+1}}{e_t}+\frac{1}{L}\left[2 \delta  \chi _A+(2-\nu) \chi _S\right].
\end{equation}

Adding the two contributions leads to
\begin{align}
&W^\text{nonloc} = -\frac{\pi  \nu^2}{30 e_t^3 a_r^{7/2}} \left[74 \ln \left(4a_r\right)-315\right] \nonumber\\
&\quad
+ \frac{\pi \nu^2 }{60 e_t^4 a_r^5} \Big\lbrace
\chi _S \left[4183 \nu -6239 + 2 (713-457 \nu ) \ln (4 a_r)\right] 
\nonumber\\
&\quad\qquad
+\delta  \chi _A \left[-6239+1426 \ln \left(4a_r\right)\right] 
\Big\rbrace,
\end{align}
where we see that $s$ cancels.
In terms of the energy and angular momentum, and expanding in $1/L$ to leading order,
\begin{align}
W^\text{nonloc} &=\frac{2\pi \nu^2 \bar{E}^2}{15 L^3} \left[315+74 \ln \left(\frac{\bar{E}}{2}\right)\right] \\
&\quad
+ \frac{2\pi \nu^2 \bar{E}^3 }{15 L^4} \bigg\lbrace\!
(4183 \nu -6239) \chi _S-6239 \delta  \chi _A \nonumber\\
&\qquad
- 2\left[713 \delta  \chi _A+(713-457 \nu ) \chi _S\right]\ln \left(\frac{E}{2}\right) 
\bigg\rbrace,\nonumber
\end{align}
and the nonlocal part of the scattering angle
\begin{align}
\chi^\text{nonloc} &= -\frac{2\pi \nu \bar{E}^2}{5 L^4} \left[315+74 \ln \left(\frac{\bar{E}}{2}\right)\right] \\
&\quad
- \frac{8\pi \nu \bar{E}^3 }{15 L^5} \bigg\lbrace\!
(4183 \nu -6239) \chi _S-6239 \delta  \chi _A \nonumber\\
&\qquad
- 2\left[713 \delta  \chi _A+(713-457 \nu ) \chi _S\right]\ln \left(\frac{E}{2}\right) 
\bigg\rbrace.\nonumber
\end{align}
In terms of $b$ and $v$, using Eqs.~\eqref{Lcov} and~\eqref{Egamma},
\begin{align}
\label{chinonloc}
\chi^\text{nonloc} &= -\frac{\pi \nu}{10 b^4} \left[148 \ln \left(\frac{v}{2}\right)+315\right] \nonumber\\
&\quad
- \frac{\pi \nu v}{15 b^5} \bigg\lbrace
4 \ln \left(\frac{v}{2}\right) \left[(679 \nu -824) \chi _S-824 \delta  \chi _A\right] \nonumber\\
&\qquad
+ (6073 \nu -7184) \chi _S-7184 \delta  \chi _A
\bigg\rbrace.
\end{align}

\section{Gauge-invariant quantities for bound orbits}
\label{sec:radAction}
In this section, we obtain two gauge-invariant quantities that characterize bound orbits:  the radial action as a function of the energy and angular momentum, and the binding energy for circular orbits as a function of the orbital frequency.

\subsection{Radial action}
\label{sec:Ir}
The radial action function contains the same gauge-invariant information as the Hamiltonian, and from it several other functions can be derived that describe bound orbits, such as the periastron advance, which can be directly related to the scattering angle via analytic continuation~\cite{Kalin:2019rwq,Kalin:2019inp}. This means that the entire calculation in Sec.~\ref{sec:local} could be performed using the radial action instead of the Hamiltonian, as was done in Ref.~\cite{Antonelli:2020ybz}.

The radial action is defined by the integral
\begin{equation}
\label{IrInteg}
I_r = \frac{1}{2\pi} \oint p_r dr,
\end{equation}
and we split it into a local contribution and a nonlocal one, such that
\begin{equation}
I_r = I_r^\text{loc} + I_r^\text{nonloc}.
\end{equation}

We calculate the local part from the local EOB Hamiltonian, i.e., Eq.~\eqref{Heob} with the nonlocal parts of the potentials and gyro-gravitomagnetic factors set to zero. We invert the local Hamiltonian iteratively to obtain $p_r(\varepsilon,L,r)$ in a PN expansion, where we recall that
\begin{align}
H_\text{EOB} &= \frac{1}{\nu} \sqrt{1 + 2\nu (\gamma -1)}, \nonumber\\
\varepsilon &\equiv \gamma^2 - 1,
\end{align}
with $\varepsilon<0,\, \gamma < 1$ for bound orbits.
Then, we integrate
\begin{equation}
I_r = \frac{1}{\pi} \int_{r_-}^{r_+} p_r(\varepsilon,L,r) \, dr,
\end{equation}
where $r_\pm$ are the zeros of the Newtonian-order $p_r^{(0)} = \sqrt{\varepsilon + 2/r - L^2/r^2}$, which are given by
\begin{equation}
r_\pm = \frac{1 \pm \sqrt{1 + L^2\varepsilon}}{-\varepsilon}.
\end{equation}
It is convenient to express the radial action in terms of the \emph{covariant} angular momentum $L_\text{cov}=L-\Delta L$, with $\Delta L$ given by Eq.~\eqref{Lcov}, since it can then be directly related to the coefficients of the scattering angle, as discussed in Ref.~\cite{Antonelli:2020ybz}, and leads to slightly simpler coefficients for the SO part.

We obtain for the local part
\begin{align}
I_r^\text{loc} &= -L + I_0
+ \frac{I_1}{\Gamma L_\text{cov}} + \frac{I_2^s}{(\Gamma L_\text{cov})^2} \\
&\quad + \frac{I_3}{(\Gamma L_\text{cov})^3} + \frac{I_4^s}{(\Gamma L_\text{cov})^4} 
+ \frac{I_5}{(\Gamma L_\text{cov})^5} + \frac{I_6^s}{(\Gamma L_\text{cov})^6} \nonumber\\
&\quad + \frac{I_7}{(\Gamma L_\text{cov})^7} + \frac{I_8^s}{(\Gamma L_\text{cov})^8} 
+ \frac{I_9}{(\Gamma L_\text{cov})^9} + \frac{I_{10}^s}{(\Gamma L_\text{cov})^{10}},\nonumber
\end{align}
where each term starts at a given PN order, with 0.5PN order corresponding to each power in $1/L$. Also, as noted in Ref.~\cite{Bini:2020wpo}, when the radial action is written in this form, in terms of $\Gamma$, the coefficients $I_{2n+1}^{(s)}$ become simple polynomials in $\nu$ of degree $\lfloor n \rfloor$.

The coefficients $I_n$ for the nonspinning local radial action up to 5PN order are given by Eq.~(13.20) of Ref.~\cite{Bini:2020wpo}.
The SO coefficients $I_n^s$ were derived in Ref.~\cite{Antonelli:2020ybz} to the 4.5PN order, but we list them here for completeness.
The coefficients $I_0,I_1,I_2^s$ are exact, and are given by
\begin{align}
I_0 &= \frac{1 + 2\varepsilon}{\sqrt{-\varepsilon}}, \nonumber\\
I_1 &= \frac{3}{4} (4 + 5\varepsilon), \nonumber\\
I_2^s &=- \frac{1}{4} \gamma  (5 \varepsilon +2) \left(4 a_b+3 a_t\right),
\end{align}
where $a_b\equiv S/M, a_t \equiv S^*/M$. The other SO coefficients, up to 5.5PN, read
\begin{widetext}
\begin{align}
I_4^s &= -\frac{21}{64} \gamma  \left(33 \varepsilon ^2+36 \varepsilon +8\right) \left(8 a_b+5 a_t\right) 
+ \gamma \nu \bigg\lbrace
\frac{21 a_b}{8}+\frac{9 a_t}{4} + \varepsilon  \left(\frac{495 a_b}{16}+\frac{219 a_t}{8}\right)
+\varepsilon ^2 \bigg[\left(\frac{17423}{192}-\frac{241 \pi ^2}{512}\right) a_b \nonumber\\
&\quad\qquad
+\left(\frac{2759}{32}-\frac{123 \pi ^2}{128}\right) a_t\bigg]
-\varepsilon ^3 \left[\left(\frac{156133}{3200}-\frac{22301 \pi ^2}{4096}\right) a_b+\left(\frac{8381 \pi ^2}{16384}-\frac{6527}{960}\right) a_t\right] + \Order(\varepsilon^4)
\bigg\rbrace, \nonumber\\
I_6^s &= \left(-\frac{25 \nu ^2}{8}+\frac{1755 \nu }{16}-\frac{495}{2}\right) a_b+\left(-\frac{45 \nu ^2}{16}+\frac{165 \nu }{2}-\frac{1155}{8}\right) a_t \nonumber\\
&\quad
-\varepsilon  \bigg\lbrace\left[\frac{645 \nu ^2}{8}+\left(\frac{3665 \pi ^2}{256}-\frac{39715}{24}\right) \nu +\frac{3465}{2}\right] a_b 
+\left[\frac{1185 \nu ^2}{16}+\left(\frac{1845 \pi ^2}{128}-\frac{10305}{8}\right) \nu +\frac{8085}{8}\right] a_t\bigg\rbrace \nonumber\\
&\quad
+\varepsilon ^2 \bigg\lbrace a_b \left(\left(\frac{10640477}{1920}-\frac{176785 \pi ^2}{2048}\right) \nu +\nu ^2 \left(\frac{45 \ms X_{59}^{\nu^2}}{128}+\frac{2755 \pi ^2}{512}-\frac{176815}{384}\right)-\frac{121275}{32}\right) \nonumber\\
&\quad\qquad
+a_t \left(\left(\frac{433715}{96}-\frac{1748755 \pi ^2}{16384}\right) \nu +\nu ^2 \left(\frac{45 \ms X_{59}^{\nu^2}}{128}+\frac{5535 \pi ^2}{1024}-\frac{26175}{64}\right)-\frac{282975}{128}\right)\bigg\rbrace + \Order(\varepsilon^3), \nonumber\\
I_8^s &=  \left[\frac{455 \nu ^3}{128}-\frac{10185 \nu ^2}{32}+\left(\frac{3755465}{1152}-\frac{42875 \pi ^2}{1536}\right) \nu -\frac{25025}{8}\right] a_b \nonumber\\
&\quad
+\left[\frac{105 \nu ^3}{32}-\frac{16485 \nu ^2}{64}+\left(\frac{437605}{192}-\frac{1435 \pi ^2}{64}\right) \nu -\frac{225225}{128}\right] a_t \nonumber\\
&\quad
+\varepsilon  \bigg\lbrace
a_b \left[\frac{4935 \nu ^3}{32}+\left(\frac{8263591}{180}-\frac{9948785 \pi ^2}{12288}\right) \nu +\nu ^2 \left(\frac{105 \ms X_{59}^{\nu^2}}{64}-\frac{1583995}{192}+\frac{27895 \pi ^2}{256}\right)-\frac{225225}{8}\right] \nonumber\\
&\quad\qquad
+a_t \left[\frac{1155 \nu ^3}{8}+\left(\frac{4594121}{144}-\frac{29957165 \pi ^2}{49152}\right) \nu +\nu ^2 \left(\frac{105 \ms X_{59}^{\nu^2}}{64}-\frac{209195}{32}+\frac{47355 \pi ^2}{512}\right)-\frac{2027025}{128}\right]\bigg\rbrace
+ \Order(\varepsilon^2), \nonumber\\
I_{10}^s &= a_b \left[-\frac{63 \nu ^4}{16}+\frac{90405 \nu ^3}{128}+\left(\frac{88995311}{1280}-\frac{545853 \pi ^2}{512}\right) \nu +\nu ^2 \left(\frac{189 \ms X_{59}^{\nu^2}}{128}+\frac{109515 \pi ^2}{512}-\frac{1119461}{64}\right)-\frac{1322685}{32}\right] \nonumber\\
&\quad
+a_t \left[-\frac{945 \nu ^4}{256}+\frac{38115 \nu ^3}{64}+\left(\frac{14456349}{320}-\frac{11632089 \pi ^2}{16384}\right) \nu +\nu ^2 \left(\frac{189 \ms X_{59}^{\nu^2}}{128}+\frac{167895 \pi ^2}{1024}-\frac{3292149}{256}\right)-\frac{2909907}{128}\right] \nonumber\\
&\quad + \Order(\varepsilon).
\end{align}
\end{widetext}

The nonlocal part can be calculated similarly by starting from the total Hamiltonian, expanding Eq.~\eqref{IrInteg} in eccentricity, then subtracting the local part.
Alternatively, it can be calculated directly from the nonlocal Hamiltonian via~\cite{Bini:2020hmy}
\begin{equation}
I_r^\text{nonloc} = - \frac{H_\text{nonloc}}{\Omega_r} ,
\end{equation}
where $\Omega_r=2\pi/T_r$ is the radial frequency given by Eq.~\eqref{nExp}.

The nonlocal Hamiltonian $H_\text{nonloc}$ in Eq.~\eqref{Hnonloc} is expressed in terms of $(e_t,a_r)$, but we can use Eqs.~\eqref{eaEL} and \eqref{eret} to obtain $I_r^\text{nonloc}(E,L)$, i.e., as a function of energy and angular momentum. 
Then, we replace $E$ with $(e_t,L)$ using Eq.~\eqref{etEL}, expand in eccentricity to $\Order(e_t^8)$, and revert back to $(E,L)$. This way, we obtain an expression for $I_r^\text{nonloc}$ in powers of $1/L$ that is valid to eighth order in eccentricity, and in which each $\varepsilon^n$ contributes up to order $e^{2n}$.

The result for the 4PN and 5.5PN SO contributions reads
\begin{widetext}
\begin{align}
\frac{I_r^\text{nonloc}}{\nu} &= \frac{1}{L_\text{cov}^7} \left(\frac{170 \ln L_\text{cov}}{3}-\frac{170 \gamma_E }{3}+\frac{18299}{96}-\frac{4777903}{90}  \ln 2+\frac{13671875 \ln 5}{512}-\frac{15081309 \ln 3}{2560}\right) \nonumber\\
&\quad
+\frac{\varepsilon}{L_{\text{cov}}^5}  \left(\frac{244 \ln L_\text{cov}}{5}-\frac{244 \gamma_E }{5}+\frac{157823}{360}-\frac{10040414}{45}  \ln 2+\frac{126953125 \ln 5}{1152}-\frac{13542147 \ln 3}{640}\right) \nonumber\\
&\quad 
+ \frac{\varepsilon ^2}{L_{\text{cov}}^3} \left(\frac{74 \ln L_\text{cov}}{15}-\frac{74 \gamma_E }{15}+\frac{89881}{240}-\frac{5292281}{15}  \ln 2+\frac{130859375 \ln 5}{768}-\frac{35029179 \ln 3}{1280}\right) \nonumber\\
&\quad 
+\frac{\varepsilon ^3}{L_{\text{cov}}}  \left(\frac{6187}{40}-\frac{11186786}{45}  \ln 2+\frac{44921875 \ln 5}{384}-\frac{1878147 \ln 3}{128}\right)\nonumber\\
&\quad
+\varepsilon ^4 L_{\text{cov}} \left(\frac{40253}{1440}-\frac{1185023}{18}  \ln 2+\frac{138671875 \ln 5}{4608}-\frac{6591861 \ln 3}{2560}\right) \nonumber\\
&\quad
+ a_b \bigg[
\frac{1}{L_{\text{cov}}^{10}} \left(-\frac{12579 \ln L_\text{cov}}{5}-\frac{24068101}{2880}+\frac{12579 \gamma_E }{5}+\frac{398742736 \ln 2}{135}+\frac{281496303 \ln 3}{1024}-\frac{40131484375 \ln 5}{27648}\right) \nonumber\\
&\qquad
+\frac{\varepsilon }{L_{\text{cov}}^8} \left(-2499 \ln L_\text{cov}-\frac{13345921}{720}+2499 \gamma_E +\frac{1548980449 \ln 2}{135}+\frac{1132984827 \ln 3}{1280}-\frac{38230234375 \ln 5}{6912}\right) \nonumber\\
&\qquad
+\frac{\varepsilon ^2}{L_{\text{cov}}^6} \left(-537 \ln L_\text{cov}-\frac{7268749}{480}+537 \gamma_E +\frac{149780983 \ln 2}{9}+\frac{2572548417 \ln 3}{2560}-\frac{36141484375 \ln 5}{4608}\right)\nonumber\\
&\qquad
+\frac{\varepsilon ^3}{L_{\text{cov}}^4} \left(-13 \ln L_\text{cov}-\frac{4143337}{720}+13 \gamma_E +\frac{1439288647 \ln 2}{135}+\frac{116812287 \ln 3}{256}-\frac{33865234375 \ln 5}{6912}\right)\nonumber\\
&\qquad
+\frac{\varepsilon ^4}{L_{\text{cov}}^2} \left(-\frac{2608213}{2880}+\frac{342877711 \ln 2}{135}+\frac{318592683 \ln 3}{5120}-\frac{31401484375 \ln 5}{27648}\right)
\bigg] \nonumber\\
&\quad
+a_t \bigg[
\frac{1}{L_{\text{cov}}^{10}} \left(-\frac{8673 \ln L_\text{cov}}{5}-\frac{16708517}{2880}+\frac{8673 \gamma_E }{5}+\frac{582216271 \ln 2}{270}+\frac{980901819 \ln 3}{5120}-\frac{29135234375 \ln 5}{27648}\right) \nonumber\\
&\qquad
+\frac{\varepsilon }{L_{\text{cov}}^8} \left(-\frac{5593 \ln L_\text{cov}}{3}-\frac{1921829}{144}+\frac{5593 \gamma_E }{3}+\frac{231474971 \ln 2}{27}+\frac{806618331 \ln 3}{1280}-\frac{28420234375 \ln 5}{6912}\right) \nonumber\\
&\qquad
+\frac{\varepsilon ^2}{L_{\text{cov}}^6} \left(-455 \ln L_\text{cov}-\frac{5444717}{480}+455 \gamma_E +\frac{191532524 \ln 2}{15}+\frac{1877210721 \ln 3}{2560}-\frac{9203828125 \ln 5}{1536}\right) \nonumber\\
&\qquad
+\frac{\varepsilon ^3}{L_{\text{cov}}^4} \left(-\frac{69 \ln L_\text{cov}}{5}-\frac{3226241}{720}+\frac{69 \gamma_E }{5}+\frac{227762869 \ln 2}{27}+\frac{87726159 \ln 3}{256}-\frac{26708984375 \ln 5}{6912}\right) \nonumber\\
&\qquad
+\frac{\varepsilon ^4}{L_{\text{cov}}^2} \left(-\frac{2112181}{2880}+\frac{562665401 \ln 2}{270}+\frac{49433247 \ln 3}{1024}-\frac{25712734375 \ln 5}{27648}\right)
\bigg].
\end{align}
\end{widetext}

\subsection{Circular-orbit binding energy}
Here, we calculate the gauge-invariant binding energy $\bar{E}$ analytically in a PN expansion, as opposed to the numerical calculation in Sec.~\ref{sec:Eb} for the EOB binding energy.

For circular orbits and aligned spins, $\bar{E}$ can be calculated from the Hamiltonian~\eqref{Heob} by setting $p_r=0$ and perturbatively solving $\dot{p}_r=0=-\partial H/\partial r$ for the angular momentum $L(r)$. Then, solving $\Omega = \partial H / \partial L$ for $r(\Omega)$, and substituting in the Hamiltonian yields $\bar{E}$ as a function of the orbital frequency. 
It is convenient to express $\bar{E}$ in terms of the dimensionless frequency parameter $v_\Omega \equiv (M\Omega_\phi)^{1/3}$.

The nonspinning 4PN binding energy is given by Eq.~(5.5) of Ref.~\cite{Damour:2014jta}, and the 4.5PN SO part is given by Eq.~(5.11) of Ref.~\cite{Antonelli:2020ybz}.
We obtain for the 5.5PN SO part
\begin{widetext}
\begin{align}
\label{Ebind}
\bar{E}^{\text{5.5PN,SO}} &= \nu v_\Omega^{13} \bigg\{
S \bigg[-\frac{4725}{32} +\nu  \left(\frac{1411663}{640}-\frac{10325 \pi ^2}{64}+\frac{352 \gamma_E }{3}+\frac{2080 \ln 2}{9}\right) +  \frac{352}{3}\nu\ln v_\Omega
+\frac{310795 \nu ^3}{5184}+\frac{35 \nu ^4}{1458} \nonumber\\
&\qquad 
+ \nu^2 \left(\frac{5 \ms X_{59}^{\nu^2}}{8}+\frac{2425 \pi ^2}{864}-\frac{1975415}{5184}\right)\bigg]
+ S^* \bigg[
-\frac{2835}{128} +\nu  \left(\frac{126715}{144}-\frac{102355 \pi ^2}{1536}+\frac{160 \gamma_E }{3}+\frac{992 \ln 2}{9}\right) \nonumber\\
&\qquad
+ \frac{160 }{3}\nu \ln v_\Omega+\nu ^2 \left(\frac{5 \ms X_{59}^{\nu^2}}{8}-\frac{205 \pi ^2}{576}-\frac{275245}{3456}\right)
+\frac{46765 \nu ^3}{864}+\frac{875 \nu ^4}{31104}
\bigg]
\bigg\}.
\end{align}
\end{widetext}

\section{Conclusions}
\label{sec:conc}
Improving the spin description in waveform models is crucial for GW observations with the continually increasing sensitivities of the Advanced LIGO, Virgo, and KAGRA detectors~\cite{KAGRA:2013rdx}, and for future GW detectors, such as the Laser Interferometer Space Antenna (LISA)~\cite{LISA:2017pwj}, the Einstein Telescope (ET)~\cite{Punturo:2010zz}, the DECi-hertz Interferometer Gravitational wave Observatory (DECIGO)~\cite{Kawamura:2006up}, and Cosmic Explorer (CE)~\cite{Reitze:2019iox}.
More accurate waveform models can lead to better estimates for the spins of binary systems, and for the orthogonal component of spin in precessing systems, which helps in identifying their formation channels~\cite{LIGOScientific:2018jsj,LIGOScientific:2020kqk}.
For this purpose, we extended in this paper the SO coupling to the 5.5PN level. 

We employed an approach~\cite{Bini:2019nra,Antonelli:2020ybz} that combines several analytical approximation methods to obtain arbitrary-mass-ratio PN results from first-order self-force results.
We computed the nonlocal-in-time contribution to the dynamics for bound orbits in a small-eccentricity expansion, Eq.~\eqref{gyroNonloc}, and for unbound motion in a large-eccentricity expansion, Eq.~\eqref{chinonloc}.
To our knowledge, this is the first time that nonlocal contributions to the conservative dynamics have been computed in the spin sector.
For the local-in-time contribution, we exploited the simple mass-ratio dependence of the PM-expanded scattering angle and related the Hamiltonian coefficients to those of the scattering angle. 
This allowed us to determine all the unknowns at that order from first-order self-force results, except for one unknown at second order in the mass ratio, see Eqs.~\eqref{chiLoc}--\eqref{gSstrLoc}. 
We also provided the radial action, in Sec.~\ref{sec:Ir}, and the circular-orbit binding energy, in Eq.~\eqref{Ebind}, as two important gauge-invariant quantities for bound orbits.
We stress again that, although all calculations in this paper were performed for aligned spins, the SO coupling is applicable for generic precessing spins.

The local part of the 5.5PN SO coupling still has an unknown coefficient, but as we showed in Fig.~\ref{fig:Eb}, its effect on the dynamics is smaller than the difference between the 4.5 and 5.5PN orders.
Determining that unknown could be done through targeted PN calculations, as was illustrated in Ref.~\cite{Bini:2021gat}, in which the authors related the two missing coefficients at 5PN order to coefficients that can be calculated from an EFT approach.
Alternatively, one could use analytical \emph{second}-order self-force results, which might become available in the near future, given the recent work on numerically computing the binding energy and energy flux~\cite{Pound:2019lzj,Warburton:2021kwk}.
Until then, one could still use the partial 5.5PN SO results in EOB waveform models complemented by NR calibration.
Such an implementation would be straightforward, since we obtained the gyro-gravitomagnetic factors that enter directly into the \texttt{SEOBNR}~\cite{Bohe:2016gbl,Cotesta:2018fcv,Ossokine:2020kjp} and \texttt{TEOBResumS}~\cite{Nagar:2018zoe,Nagar:2019wds,Nagar:2021gss} waveform models, and less directly in the \texttt{IMRPhenom} models~\cite{Hannam:2013oca,Husa:2015iqa,Khan:2019kot,Pratten:2020fqn}, which are used in GW analyses.

\section*{Acknowledgments}
I am grateful to Alessandra Buonanno, Jan Steinhoff, and Justin Vines for fruitful discussions and for their invaluable feedback on earlier drafts of this paper. I also thank Sergei Ossokine for providing NR data for the binding energy, and thank the anonymous referee for useful suggestions.

\appendix

\section{Quasi-Keplerian parametrization}
\label{app:qkepler}

\subsection{Elliptic orbits}
\label{app:qKepBound}
For a binary in a bound orbit in the orbital plane, and using polar coordinates $(r, \phi)$, the quasi-Keplerian parametrization~\cite{damour1985general}, up to 1.5PN, reads
\begin{align}
r &= a_r (1 - e_r \cos u), \\
\ell &\equiv n t = u - e_t \sin u, \label{kepEq2}\\
\phi &= 2 K \arctan\left[\sqrt{\frac{1+e_\phi}{1-e_\phi}} \tan \frac{u}{2}\right],
\end{align}
where $a_r$ is the semi-major axis, $u$ is the eccentric anomaly, $\ell$ is the mean anomaly, $n$ is the mean motion (radial angular frequency), $K$ is the periastron advance, and ($e_r,e_t,e_\phi$) are the radial, time and phase eccentricities. Spin was included in the quasi-Keplerian parametrization in Refs.~\cite{Tessmer:2010hp,Tessmer:2012xr}. (See Fig.~2 of Ref.~\cite{Tessmer:2012xr} for a geometric picture for some of these quantities.)

The (dimensionless) harmonic-coordinates Hamiltonian with LO SO reads
\begin{equation}
H = \frac{c^2}{\nu} + \frac{p^2}{2} - \frac{1}{r} + \frac{L }{c^3r^3}\left[2 \delta  \chi_A-(\nu -2) \chi_S\right].
\end{equation}
By inserting $r = a_r (1 - e_r \cos u)$ into the Hamiltonian at periastron ($u=0$) and apastron ($u=\pi$), one can solve for the energy and angular momentum (with $p_r=0$) as a function of $a_r$ and $e_r$, i.e.,
\begin{align}
\bar{E} &= \frac{-1}{2 a_r} 
+ \frac{(\nu -2) \chi _S-2 \delta  \chi _A}{2 \sqrt{a_r^5 \left(1-e_r^2\right)}}, \\
L &= \sqrt{a_r \left(1 - e_r^2\right)}
+ \frac{\left(e_r^2+3\right)\left[2 \delta  \chi _A+(2-\nu) \chi _S\right]}{2 a_r \left(e_r^2-1\right)}, \nonumber
\end{align}
where $\bar{E} \equiv E - 1/\nu < 0$ is the dimensionless binding energy, which is negative for bound orbits, and we only include the LO nonspinning and SO terms.
These expansions can be inverted to obtain $e_r(\bar{E},L)$ and $a_r(\bar{E},L)$ leading to
\begin{align}
\label{eaEL}
e_r &=\sqrt{1+2 \bar{E} L^2} +\frac{4 E \left(1+\bar{E} L^2\right) \left[2 \delta  \chi _A+(2-\nu) \chi _S\right]}{L \sqrt{1+2 \bar{E} L^2}}, \nonumber\\
a_r &= \frac{-1}{2\bar{E}}  + \frac{2 \delta  \chi _A+(2-\nu) \chi _S}{L}.
\end{align}

The radial period $T_r$ and periastron advance $K$ can be calculated from the integrals
\begin{align}
T_r &= \oint \frac{dr}{\dot{r}} = 2 \int_{r_p}^{r_a}  \frac{dr}{\partial H/\partial p_r}, \nonumber\\
K &= \frac{1}{2\pi} \oint dr \frac{\dot{\phi}}{\dot{r}}
= 2 \int_{r_p}^{r_a} dr \frac{\partial H/\partial L}{\partial H/\partial p_r},
\end{align}
where $r_p$ and $r_a$ are the periastron and apastron separations calculated from the solution of $p_r=0$, which yields the PN expansion
\begin{align}
\label{nExp}
n &= \frac{2\pi}{T_r} = 2 \sqrt{2} (-\bar{E})^{3/2} \nonumber\\
&= \frac{1}{a_r^{3/2}} + \frac{3 \left[2 \delta  \chi _A+(2-\nu) \chi _S\right]}{2 a_r^3 \sqrt{1-e_t^2}}, \\
K &= 1 + \frac{2 (\nu -2) \chi_S-4 \delta  \chi_A}{L^3} \nonumber\\
&= 1 - \frac{4 \delta  \chi _A+2 (2-\nu) \chi _S}{a_r^{3/2} \left(1-e_t^2\right)^{3/2}}.
\end{align}

The three eccentricities ($e_r,e_t,e_\phi$) agree at LO, and can be related to each other, and to the energy and angular momentum, via
\begin{align}
e_t &= \sqrt{1+2 \bar{E} L^2} + \frac{2 \bar{E} \left[2 \delta  \chi _A+(2-\nu) \chi _S\right]}{L \sqrt{1+2 \bar{E} L^2}}, \label{etEL}\\
e_\phi &= \sqrt{1+2 \bar{E} L^2}  +\frac{4 \bar{E} \left(1+\bar{E} L^2\right) \left[2 \delta  \chi _A+(2-\nu) \chi _S\right]}{L \sqrt{1+2 \bar{E} L^2}},\label{ephiEL}\\
\frac{e_r}{e_t} &= 1 + \frac{2\bar{E}}{L} \left[2\delta \chi_A + (2-\nu) \chi_S\right], \label{eret}\\
\frac{e_\phi}{e_t} &= 1 + \frac{2\bar{E}}{L} \left[2\delta \chi_A + (2-\nu) \chi_S \right].
\label{ephiet}
\end{align}

\subsection{Hyperbolic motion}
\label{app:qKephyp}
The quasi-Keplerian parametrization for hyperbolic motion~\cite{damour1985general,Cho:2018upo} can be summarized, up to 1.5PN, by the following equations:
\begin{align}
r &= \bar{a}_r (e_r \cosh \bar{u} - 1), \\
\bar{n} t &= e_t \sinh \bar{u} - \bar{u}, \\
\phi &= 2 K \arctan\left[\sqrt{\frac{e_\phi+1}{e_\phi-1}} \tanh \frac{\bar{u}}{2}\right].
\end{align}

The equations for hyperbolic motion are related to the elliptic-orbit equations via analytic continuation from $\bar{E}<0$ to $\bar{E} > 0$, and $u \to i\bar{u}$~\cite{damour1985general}. In particular, the energy and angular momentum are given by
\begin{align}
\bar{E} &= \frac{1}{2 \bar{a}_r} 
+ \frac{(\nu -2) \chi _S-2 \delta  \chi _A}{2 \sqrt{\bar{a}_r^5 \left(e_r^2-1\right)}}, \\
L &= \sqrt{\bar{a}_r \left(e_r^2-1\right)}
- \frac{\left(e_r^2+3\right)\left[2 \delta  \chi _A+(2-\nu) \chi _S\right]}{2\bar{a}_r \left(e_r^2-1\right)}. \nonumber
\end{align}
Inverting these expansions, we obtain
\begin{align}
\bar{a}_r &= \frac{1}{2\bar{E}}  - \frac{2 \delta  \chi _A+(2-\nu) \chi _S}{L},
\end{align}
and $e_r(\bar{E},L)$ the same as in Eq.~\eqref{eaEL}.

The mean motion $\bar{n}$ and periastron advance $K$ are given by 
\begin{align}
\bar{n} &= \frac{2\pi}{T_r} = 2 \sqrt{2} \bar{E}^{3/2} \nonumber\\
&= \frac{1}{\bar{a}_r^{3/2}} - \frac{3 \left[2 \delta  \chi _A+(2-\nu) \chi _S\right]}{2 \bar{a}_r^3 \sqrt{e_t^2-1}}, \\
K &= 1 + \frac{2 (\nu -2) \chi_S-4 \delta  \chi_A}{L^3} \nonumber\\
&= 1 - \frac{4 \delta  \chi _A+2 (2-\nu) \chi _S}{\bar{a}_r^{3/2} \left(e_t^2-1\right)^{3/2}}.
\end{align}
The eccentricities $e_t$ and $e_\phi$ are given in terms of energy and angular momentum by Eqs.~\eqref{etEL} and \eqref{ephiEL}, respectively.

\bibliography{refs}

\end{document}